\documentclass[a4paper,onecolumn,11pt]{quantumarticle}
\pdfoutput=1
\usepackage[utf8]{inputenc}
\usepackage[english]{babel}
\usepackage[T1]{fontenc}
\usepackage{amsmath,amssymb,latexsym,bm}
\usepackage{braket}
\usepackage{bbm}
\usepackage{hyperref}

\usepackage{tikz}
\usepackage{lipsum}
\usepackage{svg}

\PassOptionsToPackage{compress}{natbib}

\usepackage[numbers,sort&compress]{natbib}

\usepackage{array}
\usepackage{xcolor}
\usepackage{graphicx} 
\usepackage{hhline,booktabs}
\usepackage{makecell}
\usepackage{enumitem}

\usepackage{multirow}

\newcolumntype{Y}{>{\centering\arraybackslash}X}

\newcommand{\sfigure}[2]{Figure~\hyperref[#1]{\ref{#1}(#2)}}
\newcommand{\sfigref}[2]{Fig.~\hyperref[#1]{\ref{#1}(#2)}}

\usepackage{color}
\definecolor{dkgreen}{rgb}{0,0.5,0}
\definecolor{midnightblue}{rgb}{0.39,0.58,0.93}
\definecolor{kspink}{RGB}{200,0,200}
\definecolor{appendixgreen}{RGB}{9, 145, 58}

\newcommand{\comment}[1]{}{}

\usepackage{soul}

\begin{document}

\title{Quantum sensing with critical systems: impact of symmetry, imperfections, and decoherence}

\author{Yinan Chen}
\thanks{Equal contribution}
\email{yinanc@caltech.edu}
\affiliation{
Department of Physics and Institute for Quantum Information and Matter, \\ California Institute of Technology, Pasadena, California 91125, USA}%

\author{Sara Murciano}
\thanks{Equal contribution}
\email{sara.murciano@universite-paris-saclay.fr}
\affiliation{Universit\'e Paris-Saclay, CNRS, LPTMS, 91405, Orsay, France}

\author{Pablo Sala}
\affiliation{
Department of Physics and Institute for Quantum Information and Matter, \\ California Institute of Technology, Pasadena, California 91125, USA}%
\affiliation{Walter Burke Institute for Theoretical Physics, California Institute of Technology,\\ Pasadena, California 91125, USA}
\affiliation{Department of Physics, University of California, Berkeley, CA 94720, USA}
\affiliation{Simons Institute for the Theory of Computing, University of California at Berkeley}
\author{Jason Alicea}
\affiliation{
Department of Physics and Institute for Quantum Information and Matter, \\ California Institute of Technology, Pasadena, California 91125, USA}%
\affiliation{Walter Burke Institute for Theoretical Physics, California Institute of Technology,\\ Pasadena, California 91125, USA}

\maketitle

\begin{abstract}
    
Entangled many-body states enable high-precision quantum sensing beyond the standard quantum limit. We develop interferometric sensing protocols based on quantum critical wavefunctions and compare their performance with Greenberger–Horne–Zeilinger (GHZ) and spin-squeezed states. Building on the idea of symmetries as a metrological resource~\cite{frerot2024symmetry}, we introduce a symmetry-based algorithm to identify optimal measurement strategies. We illustrate this algorithm both for magnetic systems with internal
symmetries and Rydberg-atom arrays with spatial symmetries. We study the robustness of criticality for quantum sensing under non-unitary deformations, symmetry-preserving and symmetry-breaking decoherence, and qubit loss---identifying regimes where critical systems outperform GHZ states and showing that non-unitary deformation can even enhance sensing precision. Combined with recent results on log-depth preparation of critical wavefunctions, interferometric sensing in this setting appears increasingly promising.

\end{abstract}

\tableofcontents
\section{Introduction}

Quantum sensing comprises one of the main pillars of quantum technologies, alongside quantum computation and communication. The core idea is to exploit quantum superposition and entanglement to measure physical quantities---such as gravitational, magnetic, and electric fields at atomic scales---with precision transcending classical limits \cite{Wasilewski2010,Wolfgramm2010,Aasi2013,Tse2019}. Quantum‑enhanced precision in turn enables applications in building atomic clocks \cite{Kaubruegger2021}, advancing high-resolution microscopy techniques \cite{microscopy}, and many other areas of science and technology \cite{Jun2024}.

The ultimate precision limit for the estimation of a physical quantity is set by the quantum Fisher information (QFI), a multipartite entanglement witness~\cite{pezze2009,toth2012,Hyllus2012}. In particular, the QFI quantifies how sensitively a quantum system responds to an infinitesimal change in some parameter of interest. Consider, for instance, a quantum probe prepared in a state $\rho$ dependent on a parameter $\theta$ that we wish to sense---e.g., the magnetic field in a spin system. The precision $\delta \theta$ in estimating $\theta$ satisfies the Cram\'er–Rao bound (CRB) \cite{Hauke2016}  
\begin{equation}
    \delta\theta\geq \frac{1}{\sqrt{F_Q[\rho]}},
    \label{eq:bound}
\end{equation}
where $F_Q[\rho]$ denotes the QFI (defined explicitly later in Eq.~\eqref{eq:qfi_mixed}). Equation~\eqref{eq:bound} implies that the uncertainty in estimating $\theta$ decreases as the QFI increases; consequently, the scaling of the QFI with the probe system size provides a key metric determining the optimal precision. Traditional sensors use uncorrelated (e.g., spin‑coherent) states, whose QFI scales at best linearly with the probe system size $L$, leading to the precision bound $\delta\theta\sim 1/\sqrt{L}$ known as the standard quantum limit (SQL)~\cite{Giovannetti2011}. In contrast, quantum sensors exploit multipartite entanglement and long-range correlations to surpass such classical bounds and potentially reach the Heisenberg limit \cite{giovannetti2004}, where the precision scaling improves to $\delta\theta\sim 1/L$.

Systems realizing Greenberger-Horne-Zeilinger (GHZ) states---or, more generally, any superposition of two macroscopically distinct configurations---constitute powerful probes for quantum sensing~\cite{Bollinger1996,giovannetti2004,Monz2011}, which one can understand intuitively as follows. Suppose that we want to estimate a phase by applying a small rotation uniformly to $L$ spins in a GHZ state.  Because the two branches of the superposition polarize in opposite directions, the accumulated phase difference grows linearly with the number of spins, yielding an 
$L$-fold sensitivity enhancement 
that endows GHZ states with the capacity for Heisenberg-limited sensing.  
At the same time, however, macroscopic superpositions are notoriously fragile: a single qubit loss effectively projects the system onto one branch of the cat state, destroying the relative phase and erasing the enhanced signal \cite{Ma2011}. The high susceptibility of GHZ states to the external environment has motivated
the search for more robust probes. 

A promising alternative that we focus on here are many‑body systems tuned to a quantum phase transition, which naturally exhibit quantum‑enhanced sensitivity near criticality due to long-range entanglement built into the ground state~\cite{Zanardi2006,Invernizzi2008,Tommaso2018,Montenegro2024,Agarwal2025}. A paradigmatic example is the quantum critical Ising chain---not only amenable to an array of analytical and numerical tools but also now directly realizable in modern platforms such as Rydberg‑atom arrays \cite{Bernien2017,Keesling_2019,Scholl_2023}. Unlike a GHZ state, such quantum critical wavefunctions coherently superpose polarized domains at all length scales, from isolated spin flips to clusters on the order of the system size. Under a small uniform rotation applied to all spins, each domain acquires a phase proportional to its size, and all of them contribute significantly because of the diverging correlation length. 
Moreover, since the phase accumulates across all domain sizes, a local perturbation  cannot completely erase the enhanced quantum sensing, unless it acts simultaneously on every domain scale. In this sense critical states offer more resilient probes than GHZ states. 

Dependence on the parameter $\theta$ that we aim to estimate can be embedded into the probe's quantum state in two main ways. The first approach initializes the probe into the ground state of a $\theta$-dependent Hamiltonian. Near a quantum critical point, the ground state becomes highly sensitive to small variations in $\theta$, enabling precise estimation of small shifts $\delta\theta$ that can exhibit Heisenberg-limit scaling (provided the change is implemented adiabatically)~\cite{Campos2007,Invernizzi2008,Zanardi2008,Schwandt2009,Albuquerque2010,Salvatori2014,Skotiniotis_2015,Mehboudi_2015,Mehboudi2016,George2025}. Since the probe state is in thermal equilibrium, this scheme is commonly referred to as \textit{equilibrium quantum sensing} (see also Refs.~\cite{De_Pasquale_2018,Rubio2021,sarkar2025}).

\begin{figure}
    \centering   \includegraphics[width=0.5\linewidth]{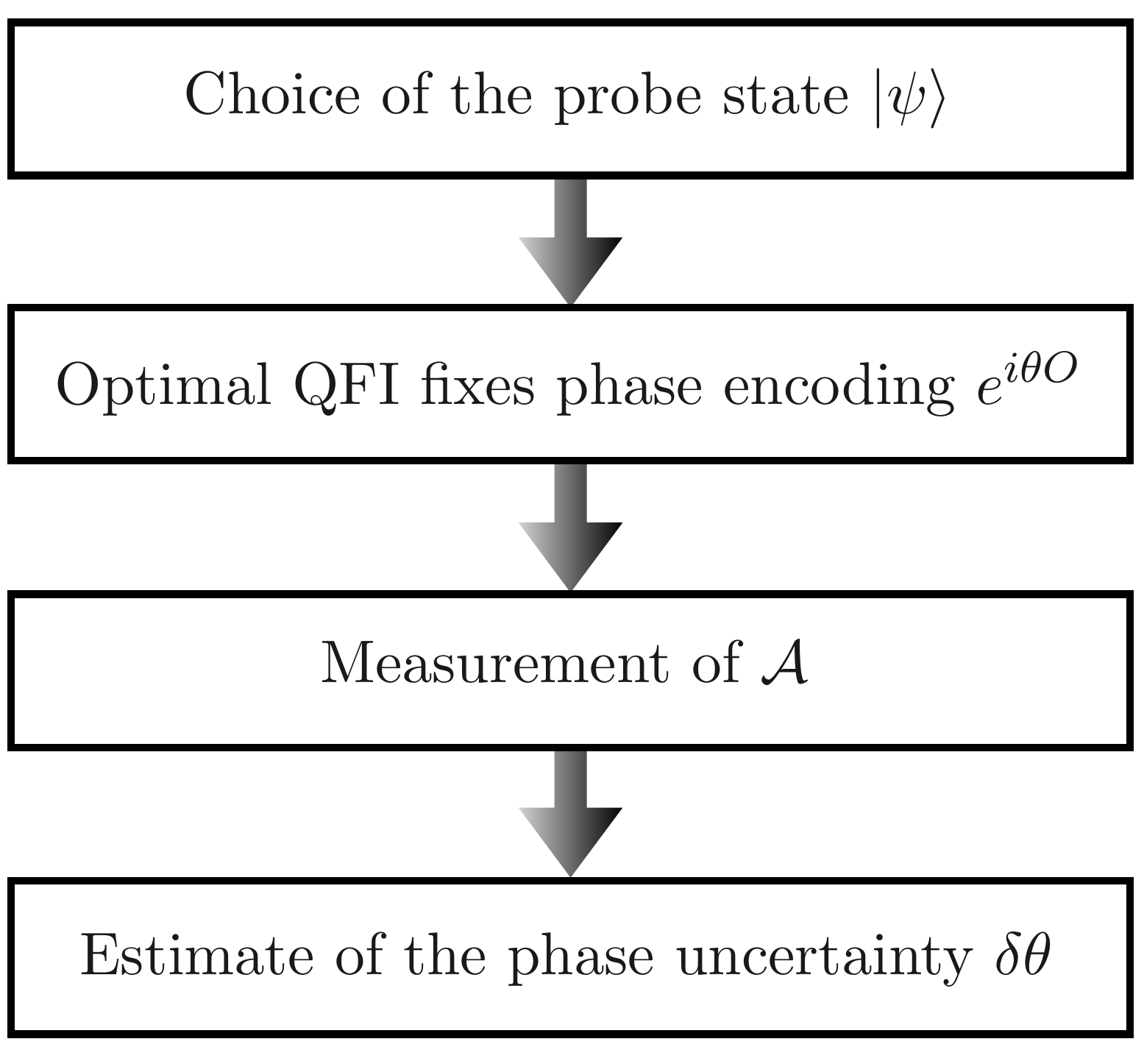}
    \caption{\textbf{Protocol.} Given a critical many-body state $\ket{\psi}$, we consider a protocol of quantum sensing in which we first encode the parameter to be estimated, $\theta$, to maximize the quantum Fisher information. An optimal measurement is then performed, yielding outcome statistics that achieve the minimal estimation uncertainty, $\delta\theta$. } 
    \label{fig:protocol}
\end{figure}

We consider the second approach wherein the quantum probe is first ``twisted’’ by the unknown parameter $\theta$ via a unitary imprinter $U(\theta)=e^{i\theta O}$, where $O$ is usually a local Hermitian operator. In the magnetic-field-sensing example, a natural choice for $O$ is the total magnetization (see footnote~\footnote{When $O$ contains $k$-body terms, in one-dimensional systems the best precision on $\delta \theta$ can even scale as $1/N^{k}$, going beyond the Heisenberg limit for $k\geq 2$ \cite{boixo07,SisiZhou2025}. However, such many-body interactions are generally more difficult to implement in practice \cite{rams2018limits}.}). After this imprinting, one reads out an observable $\mathcal{A}$, whose average value shifts in response to the $\theta$ perturbation. The observable must be chosen such that its expectation value varies significantly with respect to $\theta$, maximizing the signal (how much the observable changes with respect to variations in $\theta$) to noise (the standard deviation) ratio. Then we can quantify the uncertainty $\delta\theta$ in estimating $\theta$ using the error propagation formula, i.e., the ratio between the noise of the measurement and the signal itself. Figure~\ref{fig:protocol} summarizes the main steps of this protocol, known as \textit{interferometric quantum sensing} \cite{Rev2018,Montenegro2024,yinan2025}.

 Reference~\cite{Tommaso2018} established the utility of many-body quantum critical states for interferometric sensing based in part on universal features of their QFI \cite{Hauke2016,Giovanni2022,DiFresco2024}. For a pure probe state $\ket{\psi}$, the QFI associated with an observable $O$ is simply proportional to the variance of $O$ in that state. Thus, if $\ket{\psi}$ is a quantum critical ground state in $d$ spatial dimensions and $O = \sum_{j = 1}^L O_j$ is a sum of local operators $O_j$ with scaling dimension $\Delta$, the QFI obeys the universal scaling $\sim L^{{\rm max}[1,2(1-\Delta/d)]}$ (here $L$ denotes the total number of qubits).  The potential for sensing precision surpassing the SQL correspondingly emerges provided $\Delta < d/2$. Reference~\cite{frerot2024symmetry}  further highlighted the role that symmetries play in designing the best measurement protocol (see also Ref.~\cite{Nolan2017}). When a many-body system lies within a given symmetry sector,  correlations of charged observables are purely quantum---no matter how mixed or pure the state is---as reflected in the QFI \cite{frerot2024symmetry}. 
These correlations can then be used as a resource for precise measurements.  Our work builds on the perspective that symmetries can play a crucial role, but addresses a different and more operational question: given a probe state and a parameter-imprinting operator $O$, how can one choose a concrete readout observable $\mathcal A$ for an interferometric protocol?

Efforts have also been made to assess the robustness of systems at criticality against certain classes of decoherence: For example, even though in a context different from interferometric quantum sensing, Ref.~\cite{chen2021effects} showed that when the parameter to be estimated is encoded in the Hamiltonian, any sensitivity beyond the SQL is lost due to local dephasing. Nonetheless, it is worth noting that the combination of criticality and dissipation can sometimes lead to unexpected results. Reference~\cite{ilias2022criticality} showed that open systems admitting dissipative critical points in their steady states can exhibit enhanced QFI scaling beyond the SQL and, under suitable conditions, up to the Heisenberg limit. The robustness problem studied here is complementary to these works: we ask how noise affects an interferometric protocol performed after the probe has been prepared and the parameter has been imprinted by a unitary rotation.

Even when successfully shielding a quantum critical system from decoherence, important challenges and questions remain in using critical systems as quantum sensors. Their adiabatic preparation can lead to an inevitable growth of the protocol duration as system size increases, due to the corresponding reduction of the finite-size excitation gap.  Digital state preparation can, fortunately, proceed  more efficiently---requiring only log-$L$-depth circuits \cite{Haegeman_18,Lessa_22,daniel_25} (see Sec.~\ref{sec:conclusion} for further discussion). Once prepared with high fidelity, what is a good measurement operator to optimally estimate the imprinted parameter $\theta$ given the symmetries of the resource critical state? Can spatial symmetries play any role in systems that lack appropriate internal symmetries? Moving beyond idealities, how robust are critical states in interferometric sensing, and how do non-unitary deformation, dephasing, qubit loss, and other imperfections impact their QFI? Finally, the advantage in using critical systems for global quantum sensing (where no prior knowledge of the parameter is assumed) remains an active area of research \cite{Montenegro2021,Mukhopadhyay2024}, though we do not address that topic here.

With these questions in mind, we now summarize the  main new contributions of our work: a general symmetry-guided readout protocol, and the analysis of its robustness under non-unitary deformations and noise.
After a brief overview of essential quantum metrology concepts in Sec.~\ref{sec:qfi}, we carefully illustrate our symmetry-informed interferometric quantum-sensing protocol in Sec.~\ref{sec:symmetries} (the first key result of our paper). Given a critical state and the parameter $\theta$ to be estimated, our protocol proceeds in two steps (see again Fig.~\ref{fig:protocol}):
first, engineer the phase‐encoding to maximize the QFI, and second, identify and perform an optimal measurement whose outcome statistics minimizes the estimation uncertainty $\delta\theta$. The first step follows the standard logic of critical metrology---one chooses an imprinting operator $O$ whose correlations in the critical state lead to favorable QFI scaling. The novelty of our work is the second step. We provide a recipe for the choice of the best measurement protocol based on symmetries of the resource critical state.  We consider both internal and spatial symmetries; in turn our results hold relevance to a wide range of platforms including Rydberg atom arrays. In Sec.~\ref{sec:teleportation}, we examine whether quantum critical wavefunctions that undergo non-unitary deformation, arising for instance from weak measurements or imperfect teleportation \cite{sala2024}, can still outperform the SQL. Most interestingly, here we identify examples in which non-unitary deformation leads to slower decay of correlations~\cite{liu2024} and, consequently, a more favorable QFI scaling compared to the undeformed case.  This finding highlights a new application of the recently developed `measurement-altered quantum criticality' paradigm~\cite{garratt2023}. Next we analyze in Sec.~\ref{sec:robustness} how interferometric critical-state metrology resists various noise processes---spin-flip errors, local dephasing, and qubit losses. Crucially, critical states not only surpass the SQL in the noiseless case but also retain an advantage under realistic noise---degrading at worst back to the SQL. We stress that this analysis is distinct from previous studies of noisy equilibrium critical metrology~\cite{chen2021effects}, where the parameter is encoded directly in the Hamiltonian. Figure~\ref{fig:summ_metr} summarizes results from Sec.~\ref{sec:robustness} and highlights that noise can affect critical systems, but in a way that allows them to outperform noisy GHZ states. Finally, we conclude with a summary and outlook in Sec.~\ref{sec:conclusion}.  

\section{Quantum Fisher Information}\label{sec:qfi}
We first review some basic concepts in quantum metrology. In the standard interferometric sensing, starting with a probe state $\rho$ (pure or mixed density matrix), an unknown parameter $\theta$ is encoded via a $\theta$-dependent unitary $\rho_{\theta}=U(\theta)\rho U(\theta)^{\dagger}$, with $U(\theta)=e^{i\theta O}$. For local quantum sensing, we assume $\theta$ is unknown but fixed around a given value. From information theory \cite{degen2017quantum}, the upper bound on the knowledge of $\theta$ one can ever learn is given by the quantum Fisher information
\begin{equation}\label{eq:qfi_mixed}   F_{Q}[\rho]=2\sum_{\lambda_{i}+\lambda_{j}>0}\frac{\left(\lambda_{i}-\lambda_{j}\right)^{2}}{\lambda_{i}+\lambda_{j}}\left|\bra{i}O\ket{j}\right|^{2},
\end{equation}
where $\lambda_{i}$ and $\ket{i}$ are eigenvalues and eigenstates of the density matrix $\rho$. The QFI can in general depend on the probe state $\rho$, the operator $O$, and the parameter $\theta$.  Dependence on $\theta$ can arise, for example, if the density matrix is modified by a quantum channel $\rho_\theta = \Lambda_\theta[\rho]$. Nevertheless, in our interferometric setup, the QFI is independent of $\theta$ as Eq.~\eqref{eq:qfi_mixed} is invariant under the unitary operation $U(\theta)$. The optimal measurement, i.e., whose precision $\delta\theta$ saturates the CRB in Eq.~\eqref{eq:bound}, generally does depend on the value of $\theta$. Therefore, we assume that $\theta$ is approximately known in advance and aim to further refine its estimation using quantum metrology. Unless stated otherwise, we assume this \textit{a priori} value is centered at $\theta = 0$~\cite{Gerry2010,Shettell2020,Birrittella2021,Zhang2024}. For a general non-zero initial estimate $\theta_0$, the inverse unitary operation $U(\theta_0)^\dagger$ can be applied to realize the initialization at $\theta=0$. 

If the state $\rho$ is pure, Eq.~\eqref{eq:qfi_mixed} takes a simpler form given by the variance of the observable $O$:
\begin{equation}\label{eq:QFIpure}  F_{Q}[\rho]=4[\braket{O^2}-\braket{O}^2].
\end{equation}
The relation to precision on parameter estimation can then be understood intuitively as follows. If the probe is in a pure state that is an eigenstate of the operator $O$, the unitary merely imprints a global overall phase; correspondingly, in this limit we cannot learn anything about the parameter $\theta$. On the contrary, when the initial state has a large variance in $O$, the unitary $U(\theta)$ imprints a nontrivial change in the state---allowing measurements on the unitarily modified state to reveal $\theta$-dependent information. When $O=\sum_{\mathbf{j}} O_\mathbf{j}$, the variance reads 
\begin{equation}
    \braket{O^2}-\braket{O}^2=\sum_{\mathbf{i}\mathbf{j}}\braket{O_{{\mathbf{i}}} O_{{\mathbf{j}}}}_c,
    \label{eq:ConnectedCorr}
\end{equation}
where the subscript $c$ designates a connected correlator. Suppose now that $\rho$ corresponds to a pure $d$-dimensional critical system with a connected correlator decaying as $\braket{O_{\mathbf{i}} O_{\mathbf{j}}}_c\sim |{\mathbf{i}}-{\mathbf{j}}|^{-2\Delta}$, where $\Delta$ is the scaling dimension of the operator $O_{\mathbf{i}}$.  Scaling of $\sum_{\mathbf{i}\mathbf{j}}\braket{O_{{\mathbf{i}}} O_{{\mathbf{j}}}}_c$ can be obtained by comparing $\Delta$ with the spatial dimension $d$: If $\Delta>\frac{d}{2}$, Eq.~\eqref{eq:ConnectedCorr} is dominated by the short-distance region with $\mathbf{i}$ near $\mathbf{j}$, yielding $F_{Q}[\rho]\propto L^d$ in line with the SQL.   However, if $\Delta<\frac{d}{2}$, Eq.~\eqref{eq:ConnectedCorr} is dominated by the long-distance region with $|{\mathbf{i}}-{\mathbf{j}}|\gg 1$. This contribution yields $F_{Q}[\rho]\propto L^{2(d-\Delta)}$, which diverges with system size
faster than $L^d$. As mentioned in the introduction, this improved scaling highlights utility of critical systems for quantum sensing. The task is then to identify a measurement protocol to read out the information as precisely as possible. 

\section{Optimal measurements informed by symmetries}\label{sec:symmetries}

Which operator should one should measure to detect the sensing advantage afforded by critical systems, ideally to saturate the bound in  Eq.~\eqref{eq:bound}? In the following, we address this question for pure states prepared at
a quantum critical point separating two gapped phases classified according to either an internal symmetry (as for the transverse field Ising model) or a spatial symmetry (as for Rydberg atom chains).

\subsection{Internal symmetries}
\label{internal symmetries}
A useful way to identify such a measurement protocol is to exploit the presence of a discrete symmetry of the critical pure system. We aim to find an observable $\mathcal{A}$ whose measurement precision $\delta\theta$, given by the error propagation formula
\begin{equation}\label{eq:meansquare}
    \delta\theta=\Big |\frac{\sqrt{\mathrm{Var}_{\ket{\psi_{\theta}}}\mathcal{A}}}{\partial_{\theta}\braket{\mathcal{A}}_{\theta}}\Big|,
\end{equation}
saturates the bound given by the QFI in Eq.~\eqref{eq:bound}. Both the variance and the expectation value $\braket{\cdot}_\theta$ are computed in the pure state $\ket{\psi_\theta} = e^{i\theta O}\ket{\psi}$. The physical intuition behind Eq.~\eqref{eq:meansquare} is as follows: In the vicinity of the true (but unknown) value of $\theta$, the function $\braket{\mathcal{A}}_\theta$ is monotonic. A steeper slope of the curve $\braket{\mathcal{A}}_\theta$, corresponding to a larger derivative $\partial_\theta\braket{\mathcal{A}}_\theta$ in Eq.~\eqref{eq:meansquare}, implies greater sensitivity to small changes in $\theta$. At the same time, the “thickness’’ or spread of the curve is determined by the variance $\mathrm{Var}_{\ket{\psi_\theta}}(\mathcal{A})$. Thus, a sharper curve---i.e., one with a smaller variance---leads to a more accurate estimation. We refer the interested reader to Ref.~\cite{Tóth_2014} for further details. In Ref.~\cite{Sidhu2020}, an optimal observable was implicitly given by
\begin{equation}
\mathcal{A} = \theta\,\mathbb{I} + \frac{L_\theta}{F_Q[\rho]},
\end{equation}
where $L_\theta$ is the symmetric logarithmic derivative defined through
\begin{equation}
\partial_\theta \rho = \tfrac{1}{2}\left\{ L_\theta, \rho \right\},
\end{equation}
and $\rho = \ket{\psi_\theta}\bra{\psi_\theta}$ is the density matrix. As we show below, the construction of $\mathcal{A}$ can be significantly simplified in the presence of symmetries.

If we choose a not necessarily critical state with a discrete symmetry generated by an Abelian group 
$G$, a natural choice for the observable 
$\mathcal{A}$ is the generator of that symmetry, provided that it anticommutes with $O$, $\{\mathcal{A},O\}=0$. To see why, we can rewrite the expectation value as
\begin{equation}\label{eq:var_theta}
\braket{\mathcal{A}}_{\theta}=\braket{\psi|U(\theta)^{\dagger 2}\mathcal{A}|\psi}=s\braket{\psi|U(\theta)^{\dagger 2}|\psi},
\end{equation}
where we have assumed that $\mathcal{A}$ acts simply on $\ket{\psi}$, $\mathcal{A}\ket{\psi}=s\ket{\psi}$.
By differentiating the expression above with respect to $\theta$, and for $\theta$ close to $0$, we obtain 
\begin{equation}\label{eq:der}
\partial_{\theta}\braket{\mathcal{A}}_{\theta}= -4s\theta \mathrm{Var}(O)+\mathcal{O}(\theta^3),
\end{equation}
where the variance is now computed with respect to the original state $\ket{\psi}$. Above we implicitly used that $\braket{\psi|O|\psi} = 0$, which follows from the anticommutation relation $\left\{\mathcal{A}, O\right\} = 0$. Similarly, we can compute the variance, which in the small-$\theta$ regime reads 
\begin{equation}\label{eq:variance}
\begin{split}
\mathrm{Var}_{\ket{\psi_{\theta}}}\mathcal{A}&=\braket{\psi|U(\theta)^{\dagger 2}\mathcal{A}^2|\psi}-\braket{\psi|U(\theta)^{\dagger 2}\mathcal{A}|\psi}^2\\&=4s^2\theta^2\ \text{Var}(O)+\mathcal{O}(\theta^4).
\end{split}
\end{equation}
By combining Eqs.~\eqref{eq:der} and \eqref{eq:variance}, we obtain 
\begin{equation}\label{eq:main1}
    \delta\theta=\frac{1}{2\sqrt{\mathrm{Var}(O)}}+\mathcal{O}(\theta^2).
\end{equation}
This result means that, up to an $\mathcal{O}(\theta^2)$ error, we have found an observable whose measurement satisfies the bound $1/\sqrt{F_Q[\rho]}$, since for pure states the QFI reduces to $F_Q[\rho]=4\mathrm{Var}(O)$ (recall Eq.~\eqref{eq:QFIpure}). The bound is saturated in all spatial dimensions. To accurately estimate $\delta\theta$, we additionally seek $\mathrm{Var}(O)$ that overcomes the SQL, and so the choice of $\ket{\psi}$ as a quantum critical state plays a crucial role. We note that although dropping higher-order contributions in Eq.~\eqref{eq:main1} formally requires $\theta \ll 1/L$ such that $\mathcal{O}(\theta^3)$ terms become negligible (see similar discussions in Refs.~\cite{Shettell2020,campos2003,Zhang2024,Birrittella2020,frerot2024symmetry}), the advantage of using entangled states arises over a much wider range of $\theta$. In particular, for Ising criticality, $\delta\theta$ already exhibits an improvement over the SQL for $\theta \sim 1/\sqrt{L}$. This indicates that one can first obtain a coarse estimate of the phase using uncorrelated states (e.g., spin-coherent states), and subsequently refine the estimate by employing entangled states. We postpone a detailed discussion of this strategy to Sec.~\ref{sec:qubit_loss}. Moreover, we refer to the Appendix~\ref{app:connection} for a careful comparison with the projector-based formulation of Ref.~\cite{frerot2024symmetry} and for a discussion of the conditions under which it coincides with the present construction.

As a test-case of our findings, we consider the ground state of the one-dimensional Ising spin chain described by the Hamiltonian
\begin{equation}\label{eq:Ham}
    H=-J\sum_{j}Z_{j}Z_{j+1}-h\sum_{j}X_{j},
\end{equation}
which has a $\mathbb{Z}_2$ parity symmetry generated by $\mathcal{A} = \prod_j X_j$. Here, $Z_i$ and $X_i$ are Pauli spin operators at site $i$ of a length-$L$ chain with periodic boundary conditions, and we assume ferromagnetic interactions ($J>0$) and a positive transverse field ($h>0$). The critical point arises at $J=h$, to which we now specialize.  

A prerequisite for surpassing the SQL is identifying 2-point correlators that decay slower than $1/|i-j|$ to maximize the QFI. We remind the reader that at criticality, and for $L\gg 1$, one finds
\begin{equation}
    \braket{Z_{i}Z_{j}}\sim\frac{1}{|i-j|^{1/4}},
\end{equation}
while correlators along other orthogonal spin orientations decay faster. Therefore, the maximal QFI can be achieved by choosing $O=\sum_{j}Z_{j}$, which indeed satisfies the second condition $\{\mathcal{A},O\}=0$ to retrieve the best estimate of $\delta\theta$. In this specific example of the Ising model, we prove in Appendix~\ref{mean square error} that the mean-squared error is exactly the Fisher information \cite{Birrittella2021}, confirming that the chosen observable provides an optimal measurement strategy. 

Although the protocol as discussed above entails measuring the \emph{nonlocal} operator $\mathcal{A} = \prod_i X_i$, we stress that its expectation value can in practice be deduced from \emph{local} measurements of $X_i$ at all sites.  That is, multiplying the $X_i$ outcomes for each trial, repeating, and then averaging the results yields $\langle \mathcal{A}\rangle$ (provided measurement errors are sufficiently rare, which poses a more stringent requirement as system size increases).  

It is worth highlighting that the same sensing protocol can be applied to other critical systems. Another example is given by the XXZ spin chain 
\begin{equation}
    H_{XXZ}=\sum_{j}(X_{j}X_{j+1}+Y_{j}Y_{j+1}+\Delta Z_{j}Z_{j+1}),
\end{equation}
with $-1<\Delta\leq 1$.
In the continuum limit, this model maps to a Luttinger liquid with Luttinger parameter $K = \pi/(2(\pi - \arccos \Delta))$, and the staggered correlators of the Pauli spin operator $X$ decay as $ |i-j|^{-1/(2K)}$. Using as an imprinter $O=\sum_j (-1)^jX_j$, the sensitivity $\delta\theta$ obtained by measuring $\mathcal{A}=\prod_j Z_j$ scales like $\delta\theta\sim L^{1/(4K)-1}$.  As $\Delta\to -1$, $K$ increases and eventually the scaling reaches the Heisenberg limit. 

\subsection{Spatial symmetries}

We can build on the strategy developed in Section \ref{internal symmetries} to explore extensions where the role of discrete internal symmetries is instead played by spatial symmetries.  As a case study, we consider a Rydberg atom chain, which shares the same fundamental properties as an antiferromagnetic
Ising model with both transverse and longitudinal fields~\cite{Kevin2021,Fendley2004}. The longitudinal field explicitly breaks the spin-flip symmetry exploited in the previous subsection---necessitating a revised protocol.  

A minimal Hamiltonian describing the system reads~\cite{Fendley2004} 
\begin{equation}\label{eq:RydbergH}
H=\sum_{j}\left[\frac{\Omega}{2}(b_{j}+b_{j}^{\dagger})-\Delta n_{j}+V_{1}n_{j}n_{j+1}+V_{2}n_{j}n_{j+2}\right],
\end{equation}
where $b_{j}$ is a hard-core boson operator with occupation number $n_{j}=b_{j}^{\dagger}b_{j}$, $\Omega$ is the Rabi frequency, $\Delta$ is the detuning from resonance, and $V_{1}$ and $V_{2}$ are the nearest and second-nearest repulsive dipole interaction strengths \cite{Browaeys2020}.  We assume that $V_1$ is the dominant energy scale, imposing the Rydberg blockade constraint forbidding neighboring atoms from simultaneously entering the $n_j = 1$ Rydberg state.  
By varying $\Delta$, the system can evolve from a trivial, symmetric phase to a charge density wave phase that doubles the unit cell by hosting Rydberg excitations predominantly on every other site.  A continuous Ising transition, protected by single-site translation and bond-centered reflection symmetries, intervenes between these gapped phases. 

\begin{figure}
    \centering   \includegraphics[width=0.5\linewidth]{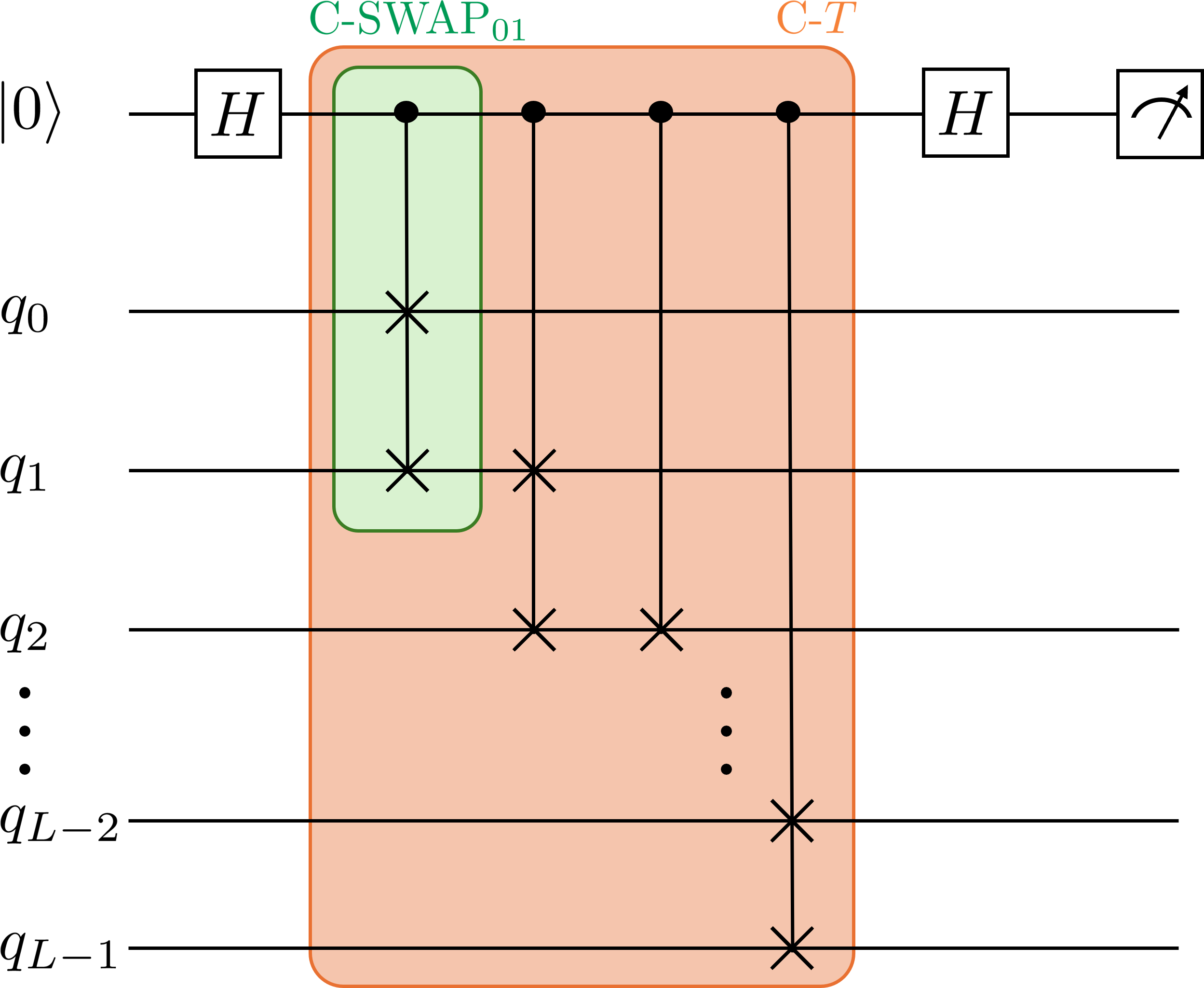}
    \caption{\textbf{A Hadamard test that measures $\mathrm{Re}\langle \psi_\theta | T | \psi_\theta \rangle$.} The orange box represents a controlled-tranlation (C-$T$) gate, which is realized by sequentially applying controlled-swap gates (green boxes) on nearest-neighbor sites. Each controlled-swap gate $\mathrm{C\text{-}SWAP}_{i,i+1}$ is decomposed into three successive Toffoli (controlled-controlled-NOT) gates, $\mathrm{C\text{-}SWAP}_{i,i+1} = \mathrm{C\text{-}CNOT}_{i,i+1}\,\mathrm{C\text{-}CNOT}_{i+1,i}\,\mathrm{C\text{-}CNOT}_{i,i+1}$. } 
    \label{fig:hadamard}
\end{figure}

A possible charge density wave order parameter is given by \cite{Kevin2021} 
\begin{align}
    \sigma_{j}&=(-1)^{j}(n_{j+1}-n_j),\label{eq:Rydberg order parameter}
\end{align}
which is the analogue of the order parameter $Z_j$ for the transverse-field Ising model.  Accordingly, at criticality one obtains $\langle \sigma_{i}\sigma_{j}\rangle\sim |i-j|^{-1/4}$. Using an imprinter with $O=\sum_j \sigma_j$ then maximizes the QFI for this model. We must also, however, identify a symmetry generator $\mathcal{A}$ that anticommutes with this choice for $O$.  For periodic chains, either the single-site translation or bond-reflection operator are candidates; for open chains, translation symmetry is absent, though at least for an even number of sites $L$, the reflection operator remains viable.   Either symmetry generator can be decomposed in terms of swap operators $S_{j,k}$ that interchange the occupation numbers at sites $j$ and $k$: $S_{j,k}\ket{n_{0}\dots n_j\dots n_k\dots n_{L-1}}=\ket{n_{1}\dots n_k\dots n_j\dots n_{L-1}}$.
Specifically, the translation operator and reflection operator (reflecting around the mid-point of bond $(j,j+1)$) respectively read
\begin{equation}\label{eq:translation}
    T=\prod_{k=0}^{L-2} S_{k,k+1},~~~~~\mathcal{I}_{j+1/2}=\prod_{k=0}^{L-2} S_{j-k,j+1+k}
\end{equation}
and satisfy $\{T,O\} = \{\mathcal{I}_{j+1/2},O\} = 0$. Since $T$ is not Hermitian ($T$ and $T^\dagger$ translate in opposite directions), it does not correspond to an observable that can be measured directly. Nevertheless, expectation values of the form $\langle \psi_\theta | T | \psi_\theta \rangle$ can be accessed experimentally using a Hadamard test (Fig.~\ref{fig:hadamard}) as follows: Starting from a $L$-qubit probe state $\ket{\psi_\theta}$ and an ancilla qubit, we sequentially apply controlled-SWAP gates $\mathrm{C\text{-}SWAP}_{i,i+1}$ to neighboring qubits $q_i$ and $q_{i+1}$, with $i = 0, \ldots, L-1$ and periodic boundary conditions $q_L = q_0$. Each controlled-SWAP gate can be decomposed into three successive Toffoli (controlled-controlled-NOT) gates according to $\mathrm{C\text{-}SWAP}_{i,i+1} = \mathrm{C\text{-}CNOT}_{i,i+1}\,\mathrm{C\text{-}CNOT}_{i+1,i}\,\mathrm{C\text{-}CNOT}_{i,i+1}$, where $\mathrm{C\text{-}CNOT}_{i,i+1}$ denotes a controlled-CNOT gate acting on qubits $q_i$ and $q_{i+1}$. A projective measurement of the ancilla qubit in the $\{\ket{0}, \ket{1}\}$ basis yields outcomes $\pm 1$ with probabilities $P_{\pm 1} = [1 \pm \mathrm{Re}\,\langle \psi_\theta | T | \psi_\theta \rangle]/2$. A direct calculation of the classical Fisher information associated with this POVM yields the desired scaling $L^{2(1-\Delta)}$ (see Appendix~\ref{mean square error}). 

A variation on the Hadamard-test circuit in Fig.~\ref{fig:hadamard} allows one to also measure expectation values $\langle \psi_\theta | \mathcal{I}_{j+1/2} | \psi_\theta \rangle$ for even-$L$ open chains where translation symmetry is absent.  (For odd-$L$ chains, we observe a similar scaling of the precision using bond-reflection, even though the operator does not anticommute with $O$). In addition, as an alternative measurement route, we note that recent work has proposed low-overhead ancilla-assisted protocols for measuring global symmetry operators using only local operations~\cite{williamson_low-overhead_2026}.

\begin{figure}
    \centering
    \includegraphics[width=0.5\linewidth]{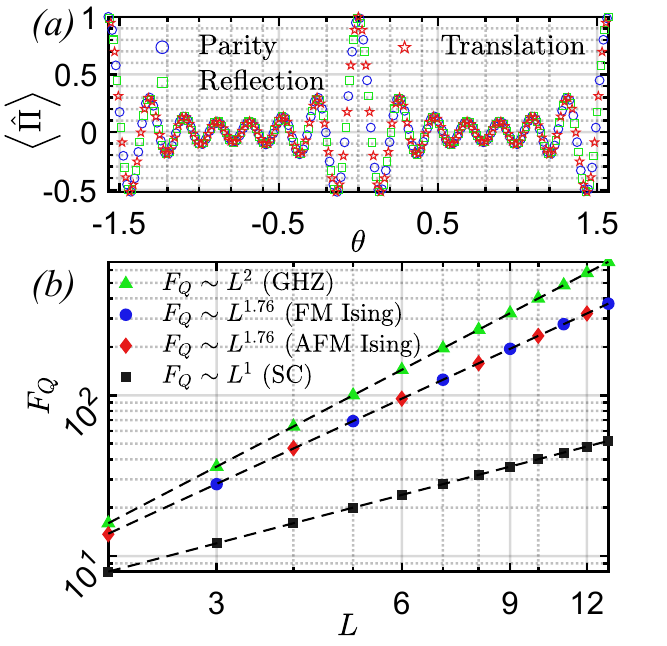}
    \caption{\textbf{Quantum Fisher information and optimal measurements.} (a) Parity, reflection and translation measurement for the Ising critical ground state at $J/h=\pm 1$. Although the simulations are done for the pure transverse-field Ising model, in the antiferromagnetic case identical scaling behavior arises if a longitudinal field $\sum_j Z_j$ is included, since it comprises an irrelevant perturbation.  (b) Quantum Fisher information $F_{Q}$ of different metrological resources as a function of system size $L$. Green triangle: GHZ state and the Heisenberg-limit scaling; Blue circle: Ising critical ground state at $J/h=1$; Red diamond: Ising critical ground state at $J/h=-1$ (only even $L$'s are considered); Dark square: Spin coherent state and the Standard-quantum-limit scaling. The $F_{Q}$'s in (b) are computed at $\theta=0$, though the QFI is independent of $\theta$ in our setup. }
    \label{QFI_Parity}
\end{figure}

To illustrate how different symmetry indicators respond to the parameter $\theta$, we examine the expectation values of several global symmetry operators as a function of $\theta$, choosing as a probe the ground state of the transverse field Ising model, Eq.~\eqref{eq:Ham}, now allowing for either sign of $J$.  
Figure~\ref{QFI_Parity}a shows that, despite originating from distinct internal or spatial symmetries, the expectation values of the parity operator (for ferromagnetic interactions) as well as the reflection and the translation operator (for antiferromagnetic interactions) all follow the same qualitative dependence on $\theta$. This demonstrates that, irrespective of which symmetry observable one chooses to measure, the resulting behavior is essentially identical. 

The possibility of finding measurement protocols in experimentally accessible platforms strengthens the relevance of the results found in this section for possible future applications. Moreover, the Hamiltonian~\eqref{eq:RydbergH} also hosts different critical phases, like the Ising tricriticality, thereby providing an ideal playground to probe multicritical behavior and a variety of universality classes in a single, tunable system.

\subsection{Comparison with non-critical optimal states}

Finally, one may ask how the results found in these sections compare to other well-known quantum states commonly employed in quantum sensing, such as the GHZ state,
\begin{equation}
\ket{\text{GHZ}} = \frac{1}{\sqrt{2}}\left(\ket{\uparrow}^{\otimes L} + \ket{\downarrow}^{\otimes L}\right).
\end{equation}
Restricting to one-dimensional systems and taking the operator $O=\sum_j Z_j$, the QFI of the pure GHZ state saturates the Heisenberg limit, scaling as $L^2$. In this case, the optimal measurement corresponds to a parity observable.
 A neat interpretation behind the fact that both GHZ and critical states surpass the SQL is that in both cases, the distribution of the eigenvalues of the operator $O=\sum_{j=1}^L Z_j$---the so-called full counting statistics---exhibits a double-peak structure that spreads out in the latter case~\cite{lamacraft} but remains sufficiently sharp that the QFI grows faster than linear in $L$. 

Similar metrological advantage can be obtained via spin-squeezed states, which are typically produced by the one-axis-twisting technique that lets spins interact so that their collective uncertainty gets squeezed along one direction \cite{Huang2024}. In this case, the QFI at the optimal twisting time scales as $L^{5/3}$, which is lower than the Heisenberg limit, but still represents a significant improvement over the SQL. Moreover, it has been shown that more sophisticated twisting protocols can restore the Heisenberg scaling \cite{Huang2024}. However, in the presence of decoherence, we will show that this optimal scaling requires changing the measurement strategy from collective spin observables to parity measurement. We explore the effects of different types of decoherence on this measurement strategy in the following sections.

Another class of states frequently used in quantum sensing are spin coherent (SC) states,
\begin{equation}
\ket{\text{SC}} = \left(\frac{\ket{\uparrow} + \ket{\downarrow}}{\sqrt{2}}\right)^{\otimes L},
\end{equation}
which are separable and exhibit classical-like behavior. Unlike the GHZ state, for the same choice of operator $O=\sum_j Z_j$, the QFI of SC states scales linearly with system size, $F_Q[\rho]\sim L$, thereby offering no advantage over the SQL. 

Figure~\ref{QFI_Parity}b summarizes the system-size scaling of the QFI, obtained from exact diagonalization.  Among the states considered, the GHZ state yields the highest QFI, fully saturating the Heisenberg limit. It is followed by ground states of the Ising spin chain with both ferromagnetic and antiferromagnetic couplings, which exhibit QFI scaling slightly below the Heisenberg limit.
Finally, the QFI of the spin coherent states grows more slowly with system size and saturates the SQL, as expected for unentangled states.

\subsection{General readout measurement}
Finally, we comment on the generality and existence of operators satisfying $\{\mathcal{A},O\}=0$. For critical points associated with symmetry-breaking transitions, the existence of a suitable symmetry operator $\mathcal{A}$ is neither rare nor fine-tuned. Such critical points are characterized by an underlying symmetry that is preserved by the Hamiltonian but acts nontrivially on the order parameter. When $O$ is chosen to be an order parameter that is odd under a $\mathbb{Z}_2$ symmetry, $\mathcal{A} O \mathcal{A}^{-1}=-O$. Thus, in these cases, the anticommutation relation follows directly from the symmetry properties of the order parameter.
\noindent The same mechanism extends beyond the specific Ising and Rydberg examples analyzed above. We have focused on spin-flip symmetry in Ising-type models and spatial symmetries in Rydberg-atom arrays because they are experimentally relevant and because the corresponding symmetry operators are naturally measurable or implementable. Beyond the critical spin models discussed above, the same anticommutation structure also appears in standard metrological probes. For GHZ and Dicke-type states, the readout can be chosen as a global parity or spin-flip operator, while the imprinting operator is a collective transverse or longitudinal spin. One-axis-twisted (OAT) spin-squeezed states provide another example, where the imprinting direction may be chosen along the anti-squeezed quadrature. Similar constructions arise in logical GHZ states encoded in quantum-error-correcting codes (QECCs), where physical Pauli operators are replaced by logical Pauli operators, and in partial-QECC settings, where the readout is determined by Pauli-$X$ support patterns with odd overlap with $Z$-type stabilizers. They can also be engineered in graph states \cite{Shettell2020}, where the compatible choices of $\mathcal A$ and $O$ are fixed by the graph structure. Finally, in two-mode bosonic probes such as NOON states, the corresponding readout is implemented by a parity measurement after a Mach-Zehnder interferometer. To make this point explicit, Tab.~\ref{tab:examples of A and O} lists possible choices of $\mathcal{A}$ and $O$ in several metrological probes and many-body systems. These examples indicate that the condition $\left\{\mathcal{A},O\right\}=0$ is widely available in many-body probes with suitable symmetries, rather than being specific to the Ising and Rydberg settings.

\begin{table}
\begin{center}
\scriptsize
\renewcommand{\arraystretch}{1.15}
\setlength{\tabcolsep}{2pt}
\begin{tabular}{|p{0.38\textwidth}|p{0.24\textwidth}|p{0.3\textwidth}|}
\hline
\textbf{System} & \textbf{Readout $A$} & \textbf{Imprinter $O$} \\
\hline
Ising FM & $\prod_j X_j$ & $\sum_j Z_j$ \\
\hline
Ising AFM &$\prod_j X_j$, $T$, $\mathcal{I}$ [Eq.~\eqref{eq:translation}] & $\sum_j(-1)^j Z_j$ \\
\hline
Ising AFM with longitudinal field (Rydberg) &$T$, $\mathcal{I}$ [Eq.~\eqref{eq:translation}] & $\sum_j(-1)^{j}(n_{j+1}-n_j)$ [Eq.~\eqref{eq:Rydberg order parameter}] \\
\hline
Tricritical Ising & $\prod_j P^X_j$ [Eq.~\eqref{eq: tricritical P operator}] & $\sum_j S^{Z}_j$ [Eq.~\eqref{eq: tricritical P operator}] \\
\hline
XXZ / Luttinger liquid \cite{musso2026} & $\prod_j Z_j$ & $\sum_j(-1)^j X_j$ \\
\hline
GHZ state & $\prod_j X_j$ & $\sum_j Z_j$ \\
\hline
Dicke probe \cite{Zou2018,musso2026} & $\prod_j Z_j$ & $\sum_j X_j$ or $\sum_j Y_j$ \\
\hline
OAT spin-squeezed state \cite{Huang2024} & $\prod_j X_j$ & $\sum_j(\cos\phi\,Y_j+\sin\phi\,Z_j)$\\
\hline
QECC probe \cite{W2014} & $\prod_j X_{L,j}$ & $\sum_j Z_{L,j}$ \\
\hline
Partial-QECC probe \cite{chen2026quantum} & $\prod_{j\in\mathcal P}X_j$ & $\sum_{j\in\mathcal S_O}G^Z_j$ \\
\hline
Bosonic probe \cite{yinan2025} & $e^{-i\pi J_X/2}(-1)^{a_1^\dagger a_1}e^{i\pi J_X/2}$ & $\frac{1}{2}(a_1^\dagger a_1-a_2^\dagger a_2)$ \\
\hline
\end{tabular}
\caption{Representative choices of the readout symmetry operator $\mathcal A$ and parameter-imprinting operator $O$ satisfying $\left\{A,O\right\}=0$ in metrological probes and many-body systems. For the tricritical Ising case, $P^X_j$ and $S_j^Z$ are defined in Eq.~\eqref{eq: tricritical P operator} for the spin-$1$ Blume-Capel model. For quantum-error-correction-code probes, we consider a logical GHZ state of $N$ logical qubits, with logical Pauli operators $X_{L,j}$ and $Z_{L,j}$. For partial quantum-error-correction-code probes, $\mathcal P$ denotes a set of Pauli-$X$ support patterns with odd overlap with each $Z$-type stabilizer generator $G^Z_j$, $j\in\mathcal S$. For two-mode bosonic probes, such as NOON states, the readout is realized by a parity measurement after a Mach-Zehnder interferometer.}
\label{tab:examples of A and O}
\end{center}
\end{table}

\section{Non-unitarily deformed critical states}\label{sec:teleportation}

So far, we have examined interferometric sensing with pristine quantum critical ground states. Next we assess the extent to which beyond-SQL precision persists when the quantum critical wavefunction used in the sensing protocol is deformed by a non-unitary operator resulting, e.g., from weakly measuring a finite fraction of the constituent degrees of freedom.   Reference~\cite{sala2024} explored a potential sensing-related scenario where this type of deformation arises: Alice prepares a pristine quantum critical state $\ket{\psi_c}$ that she wishes to transfer to Bob (who may desire to use the state for sensing but is unable to prepare it himself).  Following usual teleportation protocols, Alice and Bob begin by unitarily entangling their systems.  Alice then projectively measures her qubits and  classically communicates the outcomes to Bob, who finally applies an outcome-dependent unitary to his qubits.  If the protocol is perfect, Bob recovers precisely Alice's pristine state.  If, however, imperfections arises---for instance in the entangling unitary or measurement basis---Bob instead recovers a weakly measured counterpart of Alice's wavefunction.  Is Bob's corrupted state still metrologically useful?  Are there scenarios in which such non-unitary deformation can actually be beneficial for sensing?   

Suppose that Bob's final state takes the form
\begin{equation}\label{eq:state}
   \ket{\psi_s}\propto e^{\beta\sum_j s_j\Gamma_j}\ket{\psi_c},
\end{equation}
where $\beta$ measures the strength of the protocol imperfections, $\Gamma_j$ are Pauli operators dependent on protocol details, and $s_j \in \pm 1$ denotes Alice's measurement outcome for site $j$. We first consider the case where $\ket{\psi_c}$ is the critical ground state of the Ising chain.  For a non-unitary operator with $\Gamma_j=Z_j$ and uniform measurement outcomes where all $s_j = +1$ or all $s_j = -1$, one finds that, for any non-zero $\beta$,  $\braket{Z_j Z_k}_c$ decays as $|j-k|^{-4}$---much faster than the pristine correlator~\cite{sala2024}. Reference~\cite{sala2024} further argued that for a typical measurement outcome featuring $s_j = +1$ and $-1$ values obtained according to Born's rule, $\braket{Z_j Z_k}_c$ decays as $|j-k|^{-2}$, again much faster than the pristine case. 
This rapid decay would imply that the QFI associated with the operator $O = \sum_j Z_{j}$ asymptotically scales as $F_Q[\rho] \sim L$, thus immediately losing any quantum-criticality enhancement with this type of non-unitary deformation.  In contrast, non-unitary operators with $\Gamma_j = X_j$ or $Y_j$ deform the Ising critical state in a marginal way for uniform outcomes and an irrelevant way for typical outcomes---yielding correlations close to those of the pristine system and thereby retaining beyond-SQL QFI scaling as described in Sec.~\ref{sec:qfi}.  

\begin{figure}[t!]
    \centering
    \includegraphics[width=0.5\linewidth]{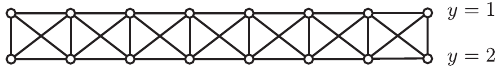}
    \caption{A generalized cluster-state with interchain-reflection symmetry constructed by completing $ZXZ$ terms on all the triangles.}
    \label{fig:cluster}
\end{figure}

Although not present in the Ising case, quantum critical states can  exhibit QFI scaling that becomes even more favorable under non-unitary deformation. We show that this scenario is possible by considering a  generalization of the  cluster-state Hamiltonian defined on a ladder geometry with interchain-reflection symmetry.  The Hamiltonian reads 
\begin{align}
    H = -\sum_{j=1}^{L-1}[ (Z_{j,1} X_{j,2} Z_{j+1,1} &+  Z_{j,2} X_{j+1,1} Z_{j+1,2}) \nonumber \\
    +(Z_{j,2} X_{j,1} Z_{j+1,2} &+ Z_{j,1} X_{j+1,2} Z_{j+1,1})],
\label{eq:SymmetricCluster}
\end{align}
where $X_{j,y}, Z_{j,y}$ are  Pauli operators acting on site $j=1,\ldots,L$ in chain $y=1,2$; see Fig.~\ref{fig:cluster} for a schematic. Reference~\cite{liu2024} showed that this model maps to two decoupled $XY$ spin chains under a non-local Kennedy-Tasaki transformation.  Moreover, adding local, symmetry-preserving terms to Eq.~\eqref{eq:SymmetricCluster} was shown to yield two decoupled Luttinger liquids characterized by the same Luttinger parameter $K$.  For the remainder of this section we assume that $\ket{\psi_c}$ is the ground state of this Luttinger liquid setup.  

In the pristine case ($\beta = 0$), when a phase is encoded on the second chain using the imprinter $U(\theta)=e^{i\theta\sum_j Z_{j,2}}$, the correlation function $\braket{Z_{j,2} Z_{k,2}}$ decays as $|j-k|^{-1/2}$ and hence admits QFI scaling as $F_Q[\rho] \sim L^{3/2}$.  This scaling already shows an enhancement relative to the SQL.
Imagine now that the wavefunction is deformed by a non-unitary operator acting on the first chain with $\Gamma_j = X_{j,1}$.  For uniform measurement outcomes with all $s_j = +1$ or all $s_j = -1$, any non-zero $\beta$ converts the Luttinger liquid into a novel `GHZ-like' state with long-range order in $\braket{Z_{j,2} Z_{k,2}}$, coexisting with residual power-law correlations \cite{liu2024}.  Thus any amount of non-unitary deformation yields asymptotic QFI scaling that saturates the Heisenberg limit!  

While the above proof-of-concept illustration specialized to a specific measurement outcome that would require post-selection, we can exploit general measurement outcomes using a decoding protocol (at least for projective measurements, $\beta \rightarrow \infty$, to which we now focus). Consider the decoded correlator 
\begin{equation}
    \braket{Z_{j,2}Z_{k,2}}_{\mathrm{d}}\equiv\sum_{s} p_{\bm{s}} \braket{{Z}_{j,2} {Z}_{k,2}}_{s} s_{j+1} s_{j+2} \cdots s_{k}
    \label{ZZd},
\end{equation}
where $p_{\bm{s}}$ is the Born probability of outcome $\bm{s}$ and $\braket{\cdot}_s$ means that the correlator is evaluated on the state~\eqref{eq:state}.  To evaluate the right-hand side, observe that the $s_j$ factors can be brought inside of the expectation value and replaced by the measured $X_{j,1}$ operators for the first chain.  The resulting string operator maps to a local expectation value under a Kennedy-Tasaki transformation, allowing one to deduce the scaling result $\braket{Z_{j,2}Z_{k,2}}_{\rm d}\sim |j-k|^{-1/(2K)}$ \cite{liu2024}.  Importantly, for $K>1$ the decoded correlator for the non-unitarily modified state decays \emph{more slowly} compared to the pristine correlator $\braket{Z_{j,2} Z_{k,2}}$.  

We can take advantage of that slower decay by defining a quantum sensing protocol where we imprint a phase on each site $j$ using  $U_j(\theta)=e^{i\theta s_1\dots s_j Z_{j,2}}$.  The associated QFI for a specific measurement outcome would read 
\begin{equation}
    F^{\bm s}_Q[\rho,\theta]=\sum_{jk}\braket{{Z}_{j,2} {Z}_{k,2}}_{s} s_{j+1} s_{j+2} \cdots s_{k}.
\end{equation}
By averaging over all measurement outcomes and using the result in Eq. \eqref{ZZd}, we get 
\begin{equation}
    F_Q[\rho]=\sum_{s} p_{\bm{s}}F^{\bm s}_Q[\rho,\theta]\sim L^{2[1-1/(4K)]}.\label{eq:aveQFI}
\end{equation}
Thus, for $K > 1$, one indeed obtains an enhancement of the QFI relative to the pristine $\beta = 0$ limit. Since the state even after non-unitary modification preserves a spin-flip symmetry on the second chain, measurement of the parity $\prod_{j}X_{j,2}$ provides an optimal readout that saturates Eq.~\eqref{eq:aveQFI}. We note that, although enhancing the averaged QFI requires an unconventional unitary imprinter $\prod_j U_j(\theta)$, it can be implemented in a straightforward manner. One first applies $X_{j,2}$ on the sites carrying a $-1$ sign in the imprinter, then applies a uniform phase imprinting $e^{i\theta \sum_j Z_{j,2}}$, and finally applies $X_{j,2}$ again on the same sites.
It is also interesting to observe that QFI scaling as in Eq.~\eqref{eq:aveQFI} could alternatively have been obtained without non-unitary deformation by imprinting the phase using a \emph{non-local} operator $O=\sum_j \prod_{i<j}X_{i,1} Z_{j,2}$. Here we designed a protocol that uses non-unitary deformation to instead utilize a set of \emph{local} operations. 

\section{Decohered critical states}\label{sec:robustness}

Our conclusions from the preceding sections relied on having a pure state, which is not what we would expect in a real setup. Here we consider mixed states and discuss sensing with quantum critical systems subject to various noise sources. We will focus primarily on the quantum critical Ising chain with local spin-flip symmetry (and without non-unitary deformation) as a concrete example. 

\subsection{Local spin flips}

We start by considering a decohered mixed state resulting from noise that preserves the $\mathbb{Z}_2$ symmetry. 
For this purpose, we consider a composition of local Pauli $X$ quantum channels that modify the density matrix via
\begin{equation}\label{eq:ZZ}
    \rho_0\to \rho =\prod_j \mathcal{E}^X_j[\rho_0].
\end{equation}
Here $\mathcal{E}^X_j(\rho_0)=(1-p)\rho_0+p X_{j}\rho_0 X_{j} $ with $\rho_0$ the initial pure critical state and $p$ the decoherence strength. While we only explicitly analyze on-site Pauli channels, we expect similar qualitative results for more general $\mathbb{Z}_2$ strongly symmetric channels.
 While, computing the QFI from the general expression in Eq.~\eqref{eq:qfi_mixed} is not trivial, symmetries still play a crucial role in streamlining the  computation. Indeed, a key observation is that $\mathcal{E}^X_j(\cdot)$ respects parity symmetry, which in turn served as the optimal measurement at $\theta\rightarrow0$.
In Appendix~\ref{app:spinflip}, we sketch the derivation of the QFI in this specific mixed state, yielding the final result
\begin{equation}\label{eq:spinflip}
    F_{Q}[\rho]=4(1-2p)^2\braket{O^2}+16p(1-p)L
\end{equation}
with $O = \sum_j Z_j$ as before and $\braket{O^2}$ evaluated in the pristine Ising model.
At the Ising critical point, as long as $p<1/2$, we recover $F_{Q}\propto L^{2(1-1/8)}$; as $p\to 1/2$, however, we should observe a sharp discontinuity to $F_{Q}=L$ in the thermodynamic limit. We corroborate our analytical result by directly comparing Eq.~\eqref{eq:spinflip} with numerical data obtained via exact diagonalization; see Fig.~\ref{Dephasing_Flip}(a). Equation~\eqref{eq:spinflip} holds also for other states respecting parity symmetry, including spin-squeezed and GHZ states; there too spin-flip channels with $p<1/2$ do not qualitatively alter the scaling with system size. This supports earlier findings in Refs.~\cite{Kurdzia2023,Kurdzia2025} that perpendicular dephasing does not destroy HL scaling.

We thus conclude that strongly parity-symmetric channels do not impact the QFI scaling, but rather only reduce its overall prefactor. Moreover, one can easily prove that our results hold also in a setup in which we apply a collection of strongly symmetric quantum channels both before and after encoding a phase through $U(\theta)$, which is described in Refs.~\cite{Demkowicz-Dobrzanski2009,Dorner2009,Zhou2018}. 
\subsection{Local dephasing}\label{sec:dephasing}

Next we consider  individual dephasing of each qubit due to an uncorrelated noise source described by quantum channels $\mathcal{E}^Z_j[\rho_0]=(1-p)\rho_{0}+pZ_{j}\rho_{0}Z_{j}$. Computing the QFI for the resulting mixed state is again far from trivial, particularly given the reduced symmetry. We therefore resort to computing the mean-square fluctuations of an observable $\mathcal{A}$, as in Eq.~\eqref{eq:meansquare}, which provides a lower bound to the QFI. Since the noise we are interested in breaks the (internal) strong $\mathbb{Z}_2$ symmetry of the system, we need to look for an observable different than the parity (otherwise $\delta\theta$ in Eq.~\eqref{eq:meansquare} would exponentially increase in system size $L$).

The goal is to find the scaling with system size of $\delta\theta$ and cross-check whether it is the same as the inverse---by means of Eq.~\eqref{eq:bound}---of the QFI, up to an overall prefactor.
Following what was done in Ref.~\cite{Chai_2025} for other potential quantum-sensing candidates, including the GHZ state, spin squeezed states, and spin coherent
states, we consider as an observable the total spin in the $y$-direction, $S_{0}=\frac{1}{2}\sum_{j}Y_{j}$. This choice is arbitrary, in the sense that we are not driven by the presence of symmetries in the system, but we only want to estimate a lower bound to the QFI. After imprinting the phase via  
$U(\theta)=e^{i\theta\sum_{j}Z_{j}}$, $S_{0}$ rotates about the $z$-axis in the $xy$-plane as $S_{\theta}= \frac{1}{2}(\cos2\theta\sum_j Y_j+\sin2\theta\sum_j X_j)$. Following the steps described in Appendix~\ref{Global and local dephasing} and applying the error propagation formula~\eqref{eq:meansquare}, we find that, close to $\theta=0$ and for $p\leq 1/2$, $\delta\theta$ is given by 
\begin{equation}\label{eq:dephasing}
    \delta\theta=\frac{\pi}{2\sqrt{L}}\sqrt{C_{y}+\frac{p(1-p)}{(1-2p)^2}}
\end{equation}
with $C_y$ a non-universal constant. This computation retrieves the SQL. The same conclusion also applies to both spin-squeezed states and spin-coherent states, although the exact form of the prefactor differs. These outcomes agree with the no-go theorem that the QFI scales at most linearly when jump operators coincide with the phase imprinter \cite{Demkowicz-Dobrzański2012}. However, for the GHZ state, local dephasing causes the estimation of 
$\delta\theta$ to become less accurate, with 
\begin{equation}\label{eq:deltaexp}
    \delta\theta=\frac{1}{2L}e^{L|\ln(1-2p)|},
\end{equation} 
growing exponentially as the system size increases. As we show in Appendix~\ref{Global and local dephasing}, a similar conclusion extends to the global dephasing channel, which directly determines the performance of the sensing procedure in the presence of collective noise. An important example is the frequency estimation for atomic clocks \cite{Kaubruegger2021,Ye2024}. We note that choosing a more general imprinting operation $U(\theta)=\exp\left(i\theta\sum_j \vec n\cdot\vec\sigma_j\right)$ does not lead to improved sensitivity $\delta\theta$ when the readout is the total spin $ S_0$. The reason is that the response $\partial_\theta\langle S_\theta\rangle$ can scale at most linearly with the system size $L$. According to the error-propagation formula~\eqref{eq:meansquare}, this implies that to maximize sensitivity, one should target observables with the slowest scaling with system size $L$ of the variance. This restricts $S_\theta$ to lie in the $xy$-plane, since $\mathrm{Var}(\sum_j Z_j)\sim L^{2(1-\Delta)}$, whereas $\mathrm{Var}(\sum_j X_j)$ and $\mathrm{Var}(\sum_j Y_j)$ scale linearly with $L$. 

\begin{figure}
    \centering
    \includegraphics[width=0.5\linewidth]{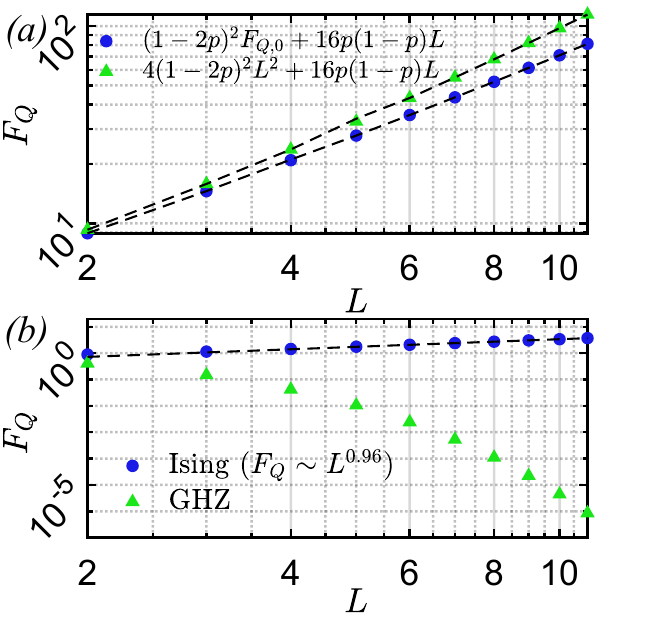}
    \caption{\textbf{Quantum Fisher information under decoherence.} (a). QFI for the Ising critical ground state (blue) and the GHZ state (green) under local bit-flip channel at $p=0.3$. Here, $F_{Q,0}$ is the QFI for the pristine Ising critical ground state at $p=0$. (b). QFI for the Ising critical ground state (blue) and the GHZ state (green) under local dephasing channel at $p=0.3$. All $F_{Q}'s$ are obtained at $\theta=0$.}
    \label{Dephasing_Flip}
\end{figure}

The result from Eq.~\eqref{eq:dephasing}, together with the bound in Eq.~\eqref{eq:bound}, provides a lower bound on the QFI, so in general one can ask whether, choosing a different measurement, the estimate $\delta \theta$ could improve.  To address this question, we evaluate the QFI numerically for small systems; Fig.~\ref{Dephasing_Flip}(b) shows its behavior under the uniform local dephasing channel. The plot confirms that, for the critical Ising spin chain (blue dots), the QFI also loses any advantage with respect to the SQL and scales linearly with system size $L$, indicating that measurements of the total spin in the $xy$-plane yield at least the optimal scaling in this regime. For spin‐squeezed states the QFI also grows only linearly with system size \cite{Chai_2025}. In contrast, when $p > 0$, the QFI of the GHZ state decays exponentially with $L$ (green triangles) 
consistent with the exponential growth of the phase uncertainty $\delta\theta$ observed in the previous example (Eq.~\eqref{eq:deltaexp}).

Therefore, even though we lose any advantage in the presence of local or global dephasing, the critical state still recovers the best possible SQL scaling, similarly to spin-squeezed states, contrary to other candidates for quantum sensing like the GHZ states, which are more fragile against these perturbations. Indeed, as we discussed in the introduction, a GHZ state relies on the superposition of two distinct configurations, and local dephasing acting on even one qubit destroys the relative phase.

\subsection{Qubit loss} \label{sec:qubit_loss}

We have shown that, in the critical Ising model, measuring the global parity operator $\Pi = \prod_j X_j$ gives optimal sensitivity to small phase shifts $\theta \to 0$, as long as the decoherence does not strongly break the system's $\mathbb{Z}_2$ symmetry. However, performing this measurement requires access to every qubit, which may not be realistic in experiments. In many setups—such as Rydberg arrays—limited control or atom loss means that some qubits cannot be addressed---effectively breaking global parity. As we have seen, losing parity symmetry can reduce the QFI back to the SQL (see Sec.~\ref{sec:dephasing}). A natural question, then, is how such imperfections affect the QFI, and whether we could still obtain enhanced sensitivity by measuring only a subregion of the system. 

To illustrate the challenge, recall that a single particle loss in a GHZ state leads to complete decoherence, resulting in a mixed state $\rho = \frac{1}{2}(\ket{0\cdots0}\bra{0\cdots0} + \ket{1\cdots1}\bra{1\cdots1})$, which has vanishing QFI with respect to $O = \sum_j Z_j$. Also, for spin squeezed states, Ref.~\cite{Li2008} shows that their robustness depends on specific correlations across all particles. If you start losing particles---whether one at a time, two at once, or three together---those correlations fall apart very quickly. We will show in this section that critical many-body states offer a more robust alternative.

We consider again the critical Ising chain in the thermodynamic limit and study $\delta\theta$ when only a contiguous subregion $[0, L_{\rm sub}]$ remains accessible. The relevant measurement is the parity operator restricted to this subregion,
\begin{equation}
    \Pi_{\rm sub}=\prod_{j=1}^{L_{\rm sub}}X_{j}.\label{Parity_sub}
\end{equation}
Note that the critical state $\ket{\psi}$ is no longer an eigenstate of $\Pi_{\rm sub}$. Accordingly, the phase can be imprinted only within the accessible subregion via the unitary $U(\theta) = e^{i O_{\rm sub} \theta}$, where $O_{\rm sub} = \frac{1}{2} \sum_{j=1}^{L_{\rm sub}} Z_j$. 

Even though we do not compute the QFI explicitly in this setting, we can still ask whether measuring the parity operator on a subsystem provides any advantage over the SQL. As we explain in detail in Appendix~\ref{appendix:XXZ}, within the interval $L_{\rm sub}^{-1}<\theta<L_{\rm sub}^{-3/4}$, the phase uncertainty $\delta\theta$ remains below the SQL. In this regime, we can identify the value of $\theta$ that maximizes the metrological sensitivity. The estimator~\eqref{eq:meansquare} shows that the phase uncertainty reaches its minimum at an intermediate angle $\theta_{\rm min}\sim L^{-7/8}_{\rm sub}$, where it scales as $\delta\theta\sim L_{\rm sub}^{-5/8}$. However, this quantum-enhanced regime presupposes that the unknown phase $\theta$
has already been localized inside the interval $[\theta_l,\theta_r]$, with $\theta_l\sim L_{\rm sub}^{-1}$ and 
$\theta_r\sim L_{\rm sub}^{-3/4}$. Locating $\theta$ within this range can be done using classical strategies with uncorrelated probes. Quantum metrology then acts as a “micrometer”: once classical methods determine the correct window, the parity measurement refines $\theta$ within it. This viewpoint suggests a practical criterion to establish whether quantum metrology offers an improvement, since the quantum uncertainty $\delta\theta$ should be smaller than the width of the initial region, $\theta_r-\theta_l$. For the Ising case considered here, $\delta\theta\sim L_{\rm sub}^{-5/8}$ is still larger than $\theta_r-\theta_l\sim L_{\rm sub}^{-3/4}$, indicating that the gain is mostly conceptual. By contrast, for the tricritical Ising universality class discussed in Appendix~\ref{appendix:XXZ}, the achievable $\delta\theta\sim L_{\rm sub}^{-31/40}$ is much finer than the corresponding window of width $L_{\rm sub}^{-13/20}$, demonstrating a clear and practically meaningful quantum advantage. In general, this analysis shows that the parity measurement within a subsystem does not always surpass the classical resolution required to identify the relevant phase window.

We conclude this section by noting that, although the parity measurement within a subsystem yields a precision surpassing the SQL, it does not saturate the ultimate bound for the Ising model. In fact, Ref.~\cite{ferro2025} has shown that the QFI of a subsystem in a critical Ising chain scales as 
$L_{\rm 
sub}^{7/4}$, matching the scaling observed for the full system. This means that measuring the parity operator on a subsystem does not saturate the Cramer-Rao bound \eqref{eq:bound}, which is $\delta\theta \geq L_{\mathrm{sub}}^{-7/8}$, while we find $\delta\theta_{\rm min}\sim L_{\rm sub}^{-5/8}$. Nevertheless, the performance of the parity measurement can be further improved for the tricritical Ising universality class, which hosts a better scaling dimension of the order parameter $\Delta_{\sigma}=\frac{3}{40}$ (cf. Appendix \ref{appendix:XXZ}).

\begin{figure*}[t!]
    \centering   \includegraphics[width=0.75\linewidth]{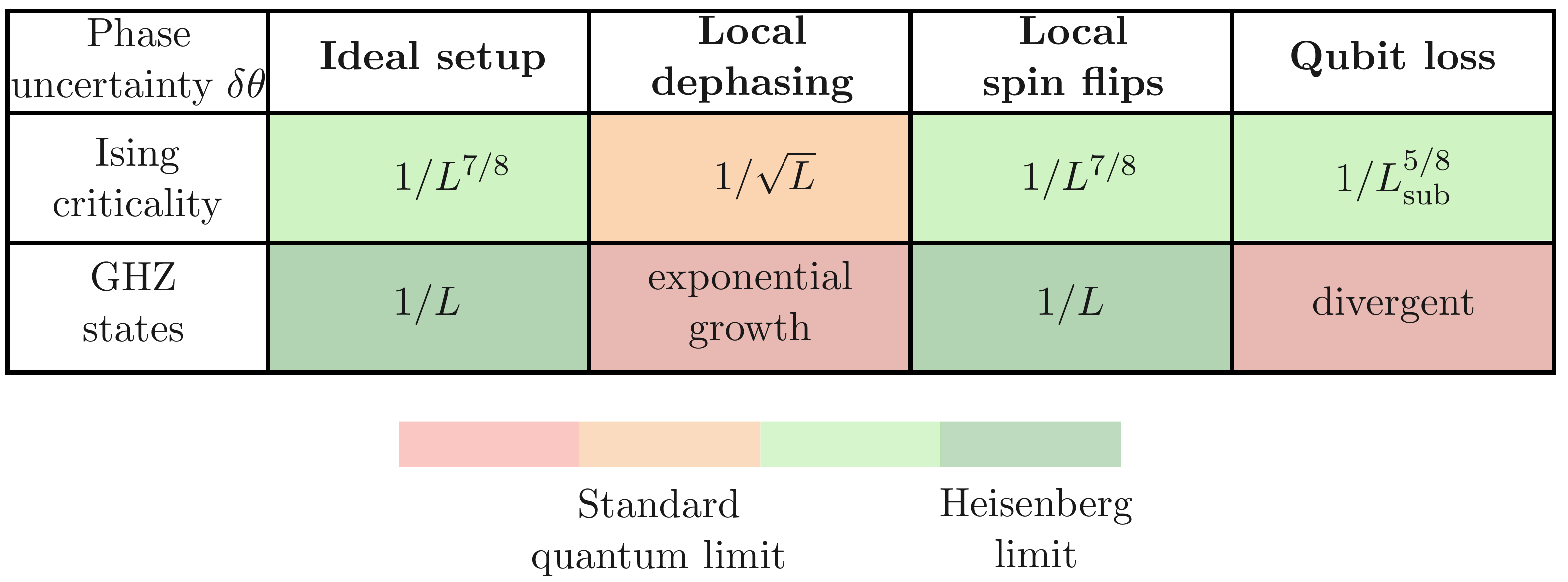}
    \caption{\textbf{Executive summary. } The table summarizes the main advantages and limitations of using a “typical’’ metrological state such as a GHZ state versus a critical wavefunction. In ideal conditions, GHZ states achieve Heisenberg scaling, while critical states interpolate between the SQL and the Heisenberg limit (e.g., for the Ising universality class, $\delta\theta\sim L^{-7/8}$). The strength of critical states emerges in the presence of decoherence, such as qubit loss, local dephasing, or spin flips, where the uncertainty at worst returns to the SQL, whereas GHZ states typically lose all metrological advantage. For OAT spin-squeezed states considered in Ref.~\cite{Kitagawa1993}, in the ideal setup $\delta\theta\sim L^{-5/6}$, while in the presence of local dephasing $\delta\theta\sim 1/\sqrt{L}$, as Ising critical states.}
    \label{fig:summ_metr}
\end{figure*}

\section{Conclusions and outlook}\label{sec:conclusion}

We have investigated various aspects of an interferometric quantum-sensing protocol, illustrated in Fig.~\ref{fig:protocol}, whose goal is to estimate with high precision a parameter $\theta$ imprinted on a critical many-body state through a unitary rotation $U(\theta)=\exp(i\theta \sum_j O_j)$. For a one-dimensional system of length $L$, the corresponding uncertainty scales as $\delta \theta\sim L^{\mathrm{max}[1,2(1-\Delta)]}$, where $\Delta$ denotes the scaling dimension of $O_j$ (see the column \textit{ideal setup} of Fig.~\ref{fig:summ_metr}). Critical systems therefore naturally interpolate between the SQL ($\delta\theta\sim L$) and the Heisenberg limit ($\delta\theta\sim L^2$). We recapitulate our main findings here:

\textbf{Optimal measurement saturating the CR bound:} Using the error-propagation formula~\eqref{eq:meansquare}, we found that the optimal observable $\mathcal{A}$ can be dictated by internal or spatial symmetries of the problem. The recipe is straightforward: (i) identify a family of generators $\sum_j O_j$ that maximizes the QFI in the critical system, and (ii) choose an observable $\mathcal{A}$ that generates a symmetry of the system and anticommutes with
$O$, i.e., $O_j$ has a well-defined charge.  This procedure ensures that $\mathcal{A}$ saturates the CR bound. We demonstrate this construction explicitly for two microscopic realizations of the Ising universality class: the Ising spin chain and Rydberg-atom arrays.

\textbf{Non-unitarily deformed wavefunctions:} We next examined whether critical wavefunctions that have been corrupted by non-unitary processes (e.g., imperfect teleportation) can still serve as useful resources for quantum sensing. We showed that the impact of this type of deformation depends on both the form of the non-unitary operator and the quantum critical wavefunction on which it acts.  In particular, the sensitivity $\delta \theta$ can either sharply diminish, remain essentially intact, or most interestingly, acquire further enhancement. We further outlined a decoding protocol in which the imprinting operation explicitly depends on the measurement outcomes and the QFI averaged over all outcomes reveals an enhancement beyond the SQL that outperforms that of the undeformed critical state.

\textbf{Effects of decoherence on critical states:} Finally, we investigated how different sources of decoherence affect metrology with Ising critical states. In the protocols considered here, the system first undergoes a quantum channel, such as local dephasing, bit flips, or qubit loss, after which the phase is imprinted. The results, summarized in Fig.~\ref{fig:summ_metr}, show a clear pattern: whenever the decoherence channel strongly breaks the $\mathbb{Z}_2$ symmetry of the system, the sensitivity $\delta\theta$ drops back to the SQL. If the symmetry is preserved, however, the decohered critical states can still achieve sensitivities between the SQL and the Heisenberg limit.

Before concluding, we also discuss the practical feasibility of exploiting critical states for quantum sensing---in particular regarding state preparation. A potential bottleneck of using critical for sensing instead of, e.g., GHZ states, is their preparation time. A natural first attempt would employ adiabatic preparation, where the time scale is fixed by the inverse gap $\sim L $.  While such adiabatic preparation intrinsically comes with complications such as populating low-energy excited states, we expect an advantage to persist provided one performs quantum sensing on a region smaller than the induced correlation length (see Section~\ref{sec:qubit_loss}). The O$(L)$ time scale for adiabatic preparation suggests that, in contrast to the O$(1)$ time required to prepare GHZ states (combining finite-depth circuits and measurements)~\cite{Zhu2023Nov,lee2022decoding}, it takes $L$-depth circuits to prepare critical states. Digital preparation of quantum critical states can, however, proceed more efficiently.  
For example, for spin models that can be mapped to free fermions, their critical ground state can be provably approximated in $\log(L)$ depth (for any spatial dimension). This result follows from either analytically constructing an entanglement renormalization scheme~\cite{Haegeman_18,Lessa_22}, or by exploiting the non-local connectivity of reconfigurable quantum systems (e.g., Rydberg atom arrays)~\cite{Gietka2022,daniel_25}

Our work highlights several open questions for future investigation. Quantifying the QFI of mixed states remains an important challenge: we were only able to derive lower bounds which, in some cases, correctly capture the scaling with system size. However, to determine whether a state truly surpasses the SQL, one needs not only the scaling with 
$L$ but also the numerical prefactor. Obtaining the full QFI, as we were able to do for the bit-flip channel, is in general very hard. It would be interesting to develop new techniques that would allow us to compute analytically or numerically the QFI in mixed states, beyond exact diagonalization. 
Moreover, our analysis has focused primarily on Ising criticality. However, Rydberg chains can realize other universality classes, such as tricritical Ising or Potts. It would be interesting to explore which symmetries become relevant in these cases and how the SQL might be surpassed. We have also shown that different types of decoherence affect the QFI in different ways. This naturally raises the question of whether one can design strategies, possibly inspired by quantum error-correction, that selectively suppress the most harmful noise sources (such as dephasing) while tolerating those that are less detrimental. It would also be interesting to investigate the potential of critical systems for global sensing tasks, and to develop explicit measurement schemes that maintain enhanced sensitivity over a broader range of phases, as suggested by our analysis of qubit-loss effects in Sec.~\ref{sec:qubit_loss}. Finally, in this work we have focused on a specific class of non-unitarily deformed states that display enhanced performance compared to pristine critical states.  We have establish this idea for a concrete model but many variations are possible. Moreover, how these deformed states respond to the different sources of decoherence discussed in Sec.~\ref{sec:robustness} remains an open question. In particular, Ref.~\cite{liu2024} identifies states in which long-range 
$Z$ order coexists with power-law correlations. Might such states be more robust to decoherence than GHZ states, or are they equally fragile?

{\bf \emph{Acknowledgments.}}~It is a pleasure to acknowledge illuminating conversations with Aash Clerk, Tuvia Gefen, Daniel González Cuadra, Yue Liu, Gil Refael, Tommaso Roscilde, and Sisi Zhou.  This work was primarily supported by the U.S. Department of Energy, Office of Science, National Quantum Information Science Research Centers, Quantum Science Center.
Additional support was provided by the Institute for Quantum Information and Matter and the Walter Burke Institute for Theoretical Physics at Caltech. P.S. acknowledges the support from the U.S. Department of Energy, Office of Science, Office of High Energy Physics, under QuantISED Award
DE-SC0019380; the NSF QLCI program through Grant No. OMA-2016245; the support from the Caltech Institute for Quantum Information and Matter, an NSF Physics
Frontiers Center (NSF Grant No.PHY-1733907), and the Walter Burke Institute for Theoretical Physics at Caltech.

\bibliographystyle{quantum}
\bibliography{reference}

@article{Nolan2017,
  title = {Optimal and Robust Quantum Metrology Using Interaction-Based Readouts},
  author = {Nolan, Samuel P. and Szigeti, Stuart S. and Haine, Simon A.},
  journal = {Phys. Rev. Lett.},
  volume = {119},
  issue = {19},
  pages = {193601},
  numpages = {7},
  year = {2017},
  month = {Nov},
  publisher = {American Physical Society},
  doi = {10.1103/PhysRevLett.119.193601},
  url = {https://link.aps.org/doi/10.1103/PhysRevLett.119.193601}
}

@article{Lessa_22,
  title = {Measurement as a Shortcut to Long-Range Entangled Quantum Matter},
  author = {Lu, Tsung-Cheng and Lessa, Leonardo A. and Kim, Isaac H. and Hsieh, Timothy H.},
  journal = {PRX Quantum},
  volume = {3},
  issue = {4},
  pages = {040337},
  numpages = {22},
  year = {2022},
  month = {Dec},
  publisher = {American Physical Society},
  doi = {10.1103/PRXQuantum.3.040337},
  url = {https://link.aps.org/doi/10.1103/PRXQuantum.3.040337}
}

@article{Haegeman_18,
  title = {Rigorous Free-Fermion Entanglement Renormalization from Wavelet Theory},
  author = {Haegeman, Jutho and Swingle, Brian and Walter, Michael and Cotler, Jordan and Evenbly, Glen and Scholz, Volkher B.},
  journal = {Phys. Rev. X},
  volume = {8},
  issue = {1},
  pages = {011003},
  numpages = {15},
  year = {2018},
  month = {Jan},
  publisher = {American Physical Society},
  doi = {10.1103/PhysRevX.8.011003},
  url = {https://link.aps.org/doi/10.1103/PhysRevX.8.011003}
}

@misc{daniel_25,
      title={Fast simulation of fermions with reconfigurable qubits}, 
      author={Nishad Maskara and Marcin Kalinowski and Daniel Gonzalez-Cuadra and Mikhail D. Lukin},
      year={2025},
      eprint={2509.08898},
      archivePrefix={arXiv},
      primaryClass={quant-ph},
      url={https://arxiv.org/abs/2509.08898}, 
}

@article{Mehboudi2016,
  title = {Achieving sub-shot-noise sensing at finite temperatures},
  author = {Mehboudi, Mohammad and Correa, Luis A. and Sanpera, Anna},
  journal = {Phys. Rev. A},
  volume = {94},
  issue = {4},
  pages = {042121},
  numpages = {7},
  year = {2016},
  month = {Oct},
  publisher = {American Physical Society},
  doi = {10.1103/PhysRevA.94.042121},
  url = {https://link.aps.org/doi/10.1103/PhysRevA.94.042121}
}

@article{Mehboudi_2015,
doi = {10.1088/1367-2630/17/5/055020},
url = {https://dx.doi.org/10.1088/1367-2630/17/5/055020},
year = {2015},
month = {may},
publisher = {IOP Publishing},
volume = {17},
number = {5},
pages = {055020},
author = {Mehboudi, M and Moreno-Cardoner, M and Chiara, G De and Sanpera, A},
title = {Thermometry precision in strongly correlated ultracold lattice gases},
journal = {New Journal of Physics}
}

@article{Salvatori2014,
  title = {Quantum metrology in Lipkin-Meshkov-Glick critical systems},
  author = {Salvatori, Giulio and Mandarino, Antonio and Paris, Matteo G. A.},
  journal = {Phys. Rev. A},
  volume = {90},
  issue = {2},
  pages = {022111},
  numpages = {11},
  year = {2014},
  month = {Aug},
  publisher = {American Physical Society},
  doi = {10.1103/PhysRevA.90.022111},
  url = {https://link.aps.org/doi/10.1103/PhysRevA.90.022111}
}

@article{Zanardi2008,
  title = {Quantum criticality as a resource for quantum estimation},
  author = {Zanardi, Paolo and Paris, Matteo G. A. and Campos Venuti, Lorenzo},
  journal = {Phys. Rev. A},
  volume = {78},
  issue = {4},
  pages = {042105},
  numpages = {7},
  year = {2008},
  month = {Oct},
  publisher = {American Physical Society},
  doi = {10.1103/PhysRevA.78.042105},
  url = {https://link.aps.org/doi/10.1103/PhysRevA.78.042105}
}

@Article{Hauke2016,
author={Hauke, Philipp
and Heyl, Markus
and Tagliacozzo, Luca
and Zoller, Peter},
title={Measuring multipartite entanglement through dynamic susceptibilities},
journal={Nature Physics},
year={2016},
month={Aug},
day={01},
volume={12},
number={8},
pages={778-782},
issn={1745-2481},
doi={10.1038/nphys3700},
url={https://doi.org/10.1038/nphys3700}
}

@article{Skotiniotis_2015,
   title={Quantum metrology for the Ising Hamiltonian with transverse magnetic field},
   volume={17},
   ISSN={1367-2630},
   url={http://dx.doi.org/10.1088/1367-2630/17/7/073032},
   DOI={10.1088/1367-2630/17/7/073032},
   number={7},
   journal={New Journal of Physics},
   publisher={IOP Publishing},
   author={Skotiniotis, Michael and Sekatski, Pavel and Dür, Wolfgang},
   year={2015},
 pages={073032} }

@article{Zhang2024,
  title = {Fast Generation of GHZ-like States Using Collective-Spin XYZ Model},
  author = {Zhang, Xuanchen and Hu, Zhiyao and Liu, Yong-Chun},
  journal = {Phys. Rev. Lett.},
  volume = {132},
  issue = {11},
  pages = {113402},
  numpages = {6},
  year = {2024},
  month = {Mar},
  publisher = {American Physical Society},
  doi = {10.1103/PhysRevLett.132.113402},
}

@article{sala2024,
  title = {Quantum Criticality Under Imperfect Teleportation},
  author = {Sala, Pablo and Murciano, Sara and Liu, Yue and Alicea, Jason},
  journal = {PRX Quantum},
  volume = {5},
  issue = {3},
  pages = {030307},
  numpages = {40},
  year = {2024},
  month = {Jul},
  publisher = {American Physical Society},
  doi = {10.1103/PRXQuantum.5.030307},
  url = {https://link.aps.org/doi/10.1103/PRXQuantum.5.030307}
}

@article{liu2024,
title = {Boundary transitions from a single round of measurements on gapless quantum states},
  author = {Liu, Yue and Murciano, Sara and Mross, David F. and Alicea, Jason},
  journal = {Phys. Rev. Res.},
  volume = {7},
  issue = {2},
  pages = {023293},
  numpages = {31},
  year = {2025},
  month = {Jun},
  publisher = {American Physical Society},
  doi = {10.1103/l4b7-h5cd},
  url = {https://link.aps.org/doi/10.1103/l4b7-h5cd}
}

@article{Yuto24,
  title = {System-environment entanglement phase transitions},
  author = {Ashida, Yuto and Furukawa, Shunsuke and Oshikawa, Masaki},
  journal = {Phys. Rev. B},
  volume = {110},
  issue = {9},
  pages = {094404},
  numpages = {20},
  year = {2024},
  month = {Sep},
  publisher = {American Physical Society},
  doi = {10.1103/PhysRevB.110.094404},
  url = {https://link.aps.org/doi/10.1103/PhysRevB.110.094404}
}

@article{rams2018limits,
  title = {At the Limits of Criticality-Based Quantum Metrology: Apparent Super-Heisenberg Scaling Revisited},
  author = {Rams, Marek M. and Sierant, Piotr and Dutta, Omyoti and Horodecki, Pawe\l{} and Zakrzewski, Jakub},
  journal = {Phys. Rev. X},
  volume = {8},
  issue = {2},
  pages = {021022},
  numpages = {16},
  year = {2018},
  month = {Apr},
  publisher = {American Physical Society},
  doi = {10.1103/PhysRevX.8.021022},
  url = {https://link.aps.org/doi/10.1103/PhysRevX.8.021022}
}

@article{ilias2022criticality,
  title = {Criticality-Enhanced Quantum Sensing via Continuous Measurement},
  author = {Ilias, Theodoros and Yang, Dayou and Huelga, Susana F. and Plenio, Martin B.},
  journal = {PRX Quantum},
  volume = {3},
  issue = {1},
  pages = {010354},
  numpages = {21},
  year = {2022},
  month = {Mar},
  publisher = {American Physical Society},
  doi = {10.1103/PRXQuantum.3.010354},
  url = {https://link.aps.org/doi/10.1103/PRXQuantum.3.010354}
}

@article{frerot2024symmetry,
  title = {Symmetry: A Fundamental Resource for Quantum Coherence and Metrology},
  author = {Fr\'erot, Ir\'en\'ee and Roscilde, Tommaso},
  journal = {Phys. Rev. Lett.},
  volume = {133},
  issue = {26},
  pages = {260402},
  numpages = {6},
  year = {2024},
  month = {Dec},
  publisher = {American Physical Society},
  doi = {10.1103/PhysRevLett.133.260402},
  url = {https://link.aps.org/doi/10.1103/PhysRevLett.133.260402}
}

@article{chen2021effects,
  title = {Effects of local decoherence on quantum critical metrology},
  author = {Chen, Chong and Wang, Ping and Liu, Ren-Bao},
  journal = {Phys. Rev. A},
  volume = {104},
  issue = {2},
  pages = {L020601},
  numpages = {5},
  year = {2021},
  month = {Aug},
  publisher = {American Physical Society},
  doi = {10.1103/PhysRevA.104.L020601},
  url = {https://link.aps.org/doi/10.1103/PhysRevA.104.L020601}
}

@article{Giovannetti2011,
author={Giovannetti, Vittorio
and Lloyd, Seth
and Maccone, Lorenzo},
title={Advances in quantum metrology},
journal={Nature Photonics},
year={2011},
month={Apr},
day={01},
volume={5},
number={4},
pages={222-229},
issn={1749-4893},
doi={10.1038/nphoton.2011.35},
url={https://doi.org/10.1038/nphoton.2011.35}
}

@article{Invernizzi2008,
  title = {Optimal quantum estimation in spin systems at criticality},
  author = {Invernizzi, Carmen and Korbman, Michael and Campos Venuti, Lorenzo and Paris, Matteo G. A.},
  journal = {Phys. Rev. A},
  volume = {78},
  issue = {4},
  pages = {042106},
  numpages = {7},
  year = {2008},
  month = {Oct},
  publisher = {American Physical Society},
  doi = {10.1103/PhysRevA.78.042106},
  url = {https://link.aps.org/doi/10.1103/PhysRevA.78.042106}
}

@article{Montenegro2024,
title = {Review: Quantum metrology and sensing with many-body systems},
journal = {Physics Reports},
volume = {1134},
pages = {1-62},
year = {2025},
note = {Review: Quantum metrology and sensing with many-body systems},
issn = {0370-1573},
doi = {https://doi.org/10.1016/j.physrep.2025.05.005},
url = {https://www.sciencedirect.com/science/article/pii/S0370157325001565},
author = {Victor Montenegro and Chiranjib Mukhopadhyay and Rozhin Yousefjani and Saubhik Sarkar and Utkarsh Mishra and Matteo G.A. Paris and Abolfazl Bayat}
}

@article{Kaubruegger2021,
  title = {Quantum Variational Optimization of Ramsey Interferometry and Atomic Clocks},
  author = {Kaubruegger, Raphael and Vasilyev, Denis V. and Schulte, Marius and Hammerer, Klemens and Zoller, Peter},
  journal = {Phys. Rev. X},
  volume = {11},
  issue = {4},
  pages = {041045},
  numpages = {21},
  year = {2021},
  month = {Dec},
  publisher = {American Physical Society},
  doi = {10.1103/PhysRevX.11.041045},
  url = {https://link.aps.org/doi/10.1103/PhysRevX.11.041045}
}

@article{microscopy,
title = {Quantum-enhanced nonlinear microscopy},
    author = {Casacio, C.A. \textit{et al.}},
    journal = {Nature},
    year = {2021},
    volume={594},
    pages={201–206},
    url={https://doi.org/10.1038/s41586-021-03528-w}
}

@article{lamacraft,
  title = {Order Parameter Statistics in the Critical Quantum Ising Chain},
  author = {Lamacraft, Austen and Fendley, Paul},
  journal = {Phys. Rev. Lett.},
  volume = {100},
  issue = {16},
  pages = {165706},
  numpages = {4},
  year = {2008},
  month = {Apr},
  publisher = {American Physical Society},
  doi = {10.1103/PhysRevLett.100.165706},
  url = {https://link.aps.org/doi/10.1103/PhysRevLett.100.165706}
}

@article{boixo07,
  title = {Generalized Limits for Single-Parameter Quantum Estimation},
  author = {Boixo, Sergio and Flammia, Steven T. and Caves, Carlton M. and Geremia, JM},
  journal = {Phys. Rev. Lett.},
  volume = {98},
  issue = {9},
  pages = {090401},
  numpages = {4},
  year = {2007},
  month = {Feb},
  publisher = {American Physical Society},
  doi = {10.1103/PhysRevLett.98.090401},
  url = {https://link.aps.org/doi/10.1103/PhysRevLett.98.090401}
}

@misc{ferro2025,
      title={Kicking Quantum Fisher Information out of Equilibrium}, 
      author={Florent Ferro and Maurizio Fagotti},
      year={2025},
      eprint={2503.21905},
      archivePrefix={arXiv},
      primaryClass={quant-ph},
      url={https://arxiv.org/abs/2503.21905}, 
}

@article{Li2008,
  title = {Optimum Spin Squeezing in Bose-Einstein Condensates with Particle Losses},
  author = {Li, Yun and Castin, Y. and Sinatra, A.},
  journal = {Phys. Rev. Lett.},
  volume = {100},
  issue = {21},
  pages = {210401},
  numpages = {4},
  year = {2008},
  month = {May},
  publisher = {American Physical Society},
  doi = {10.1103/PhysRevLett.100.210401},
  url = {https://link.aps.org/doi/10.1103/PhysRevLett.100.210401}
}

@article{Rev2018,
  title = {Quantum metrology with nonclassical states of atomic ensembles},
  author = {Pezz\`e, Luca and Smerzi, Augusto and Oberthaler, Markus K. and Schmied, Roman and Treutlein, Philipp},
  journal = {Rev. Mod. Phys.},
  volume = {90},
  issue = {3},
  pages = {035005},
  numpages = {70},
  year = {2018},
  month = {Sep},
  publisher = {American Physical Society},
  doi = {10.1103/RevModPhys.90.035005},
  url = {https://link.aps.org/doi/10.1103/RevModPhys.90.035005}
}

@Article{Demkowicz-Dobrzański2012,
author={Demkowicz-Dobrza{\'{n}}ski, Rafa{\l}
and Ko{\l}ody{\'{n}}ski, Jan
and Gu{\c{T}}{\u{a}}, M{\u{a}}d{\u{a}}lin},
title={The elusive Heisenberg limit in quantum-enhanced metrology},
journal={Nat. Commun.},
year={2012},
month={Sep},
day={18},
volume={3},
number={1},
pages={1063},
issn={2041-1723},
doi={10.1038/ncomms2067},
url={https://doi.org/10.1038/ncomms2067}
}

@article{degen2017quantum,
  title = {Quantum sensing},
  author = {Degen, C. L. and Reinhard, F. and Cappellaro, P.},
  journal = {Rev. Mod. Phys.},
  volume = {89},
  issue = {3},
  pages = {035002},
  numpages = {39},
  year = {2017},
  month = {Jul},
  publisher = {American Physical Society},
  doi = {10.1103/RevModPhys.89.035002}
}

@article{Birrittella2021,
    author = {Birrittella, Richard J. and Alsing, Paul M. and Gerry, Christopher C.},
    title = "{The parity operator: Applications in quantum metrology}",
    journal = {AVS Quantum Sci.},
    volume = {3},
    number = {1},
    pages = {014701},
    year = {2021},
    month = {03},
    issn = {2639-0213},
    doi = {10.1116/5.0026148}
}

@article{garratt2023,
  title = {Measurements Conspire Nonlocally to Restructure Critical Quantum States},
  author = {Garratt, Samuel J. and Weinstein, Zack and Altman, Ehud},
  journal = {Phys. Rev. X},
  volume = {13},
  issue = {2},
  pages = {021026},
  numpages = {24},
  year = {2023},
  month = {May},
  publisher = {American Physical Society},
  doi = {10.1103/PhysRevX.13.021026},
  url = {https://link.aps.org/doi/10.1103/PhysRevX.13.021026}
}

@article{Ye2024,
  title = {Essay: Quantum Sensing with Atomic, Molecular, and Optical Platforms for Fundamental Physics},
  author = {Ye, Jun and Zoller, Peter},
  journal = {Phys. Rev. Lett.},
  volume = {132},
  issue = {19},
  pages = {190001},
  numpages = {11},
  year = {2024},
  month = {May},
  publisher = {American Physical Society},
  doi = {10.1103/PhysRevLett.132.190001},
}

@article{Xu2023,
  title = {Quantum Criticality Under Decoherence or Weak Measurement},
  author = {Lee, Jong Yeon and Jian, Chao-Ming and Xu, Cenke},
  journal = {PRX Quantum},
  volume = {4},
  issue = {3},
  pages = {030317},
  numpages = {20},
  year = {2023},
  month = {Aug},
  publisher = {American Physical Society},
  doi = {10.1103/PRXQuantum.4.030317},
  url = {https://link.aps.org/doi/10.1103/PRXQuantum.4.030317}
}

@article{Rath2021,
  title = {Quantum Fisher Information from Randomized Measurements},
  author = {Rath, Aniket and Branciard, Cyril and Minguzzi, Anna and Vermersch, Beno\^{\i}t},
  journal = {Phys. Rev. Lett.},
  volume = {127},
  issue = {26},
  pages = {260501},
  numpages = {6},
  year = {2021},
  month = {Dec},
  publisher = {American Physical Society},
  doi = {10.1103/PhysRevLett.127.260501},
}

@article{campos2003,
  title = {Optical interferometry at the Heisenberg limit with twin Fock states and parity measurements},
  author = {Campos, R. A. and Gerry, Christopher C. and Benmoussa, A.},
  journal = {Phys. Rev. A},
  volume = {68},
  issue = {2},
  pages = {023810},
  numpages = {5},
  year = {2003},
  month = {Aug},
  publisher = {American Physical Society},
  doi = {10.1103/PhysRevA.68.023810},
  url = {https://link.aps.org/doi/10.1103/PhysRevA.68.023810}
}

@article{Birrittella2020,
    author = "Birrittella, Richard J. and Alsing, Paul M. and Gerry, Christopher C.",
    title = "{The parity operator: Applications in quantum metrology}",
    eprint = "2008.08658",
    archivePrefix = "arXiv",
    primaryClass = "quant-ph",
    doi = "10.1116/5.0026148",
    journal = "AVS Quantum Sci.",
    volume = "3",
    number = "1",
    pages = "014701",
    year = "2021"
}

@article{Kevin2021,
  title = {Microscopic characterization of Ising conformal field theory in Rydberg chains},
  author = {Slagle, Kevin and Aasen, David and Pichler, Hannes and Mong, Roger S. K. and Fendley, Paul and Chen, Xie and Endres, Manuel and Alicea, Jason},
  journal = {Phys. Rev. B},
  volume = {104},
  issue = {23},
  pages = {235109},
  numpages = {20},
  year = {2021},
  month = {Dec},
  publisher = {American Physical Society},
  doi = {10.1103/PhysRevB.104.235109},
  url = {https://link.aps.org/doi/10.1103/PhysRevB.104.235109}
}

@article{Browaeys2020,
    author = "Browaeys, Antoine and Lahaye, Thierry",
    title = "{Many-body physics with individually controlled Rydberg atoms}",
    eprint = "2002.07413",
    archivePrefix = "arXiv",
    primaryClass = "cond-mat.quant-gas",
    doi = "10.1038/s41567-019-0733-z",
    journal = "Nature Phys.",
    volume = "16",
    number = "2",
    pages = "132--142",
    year = "2020"
}

@article{Shettell2020,
  title = {Graph States as a Resource for Quantum Metrology},
  author = {Shettell, Nathan and Markham, Damian},
  journal = {Phys. Rev. Lett.},
  volume = {124},
  issue = {11},
  pages = {110502},
  numpages = {6},
  year = {2020},
  month = {Mar},
  publisher = {American Physical Society},
  doi = {10.1103/PhysRevLett.124.110502},
  url = {https://link.aps.org/doi/10.1103/PhysRevLett.124.110502}
}

@article{Gerry2010,
  title = {Heisenberg-limited interferometry with pair coherent states and parity measurements},
  author = {Gerry, Christopher C. and Mimih, Jihane},
  journal = {Phys. Rev. A},
  volume = {82},
  issue = {1},
  pages = {013831},
  numpages = {7},
  year = {2010},
  month = {Jul},
  publisher = {American Physical Society},
  doi = {10.1103/PhysRevA.82.013831},
  url = {https://link.aps.org/doi/10.1103/PhysRevA.82.013831}
}

@article{Demkowicz-Dobrzanski2009,
  title = {Quantum phase estimation with lossy interferometers},
  author = {Demkowicz-Dobrzanski, R. and Dorner, U. and Smith, B. J. and Lundeen, J. S. and Wasilewski, W. and Banaszek, K. and Walmsley, I. A.},
  journal = {Phys. Rev. A},
  volume = {80},
  issue = {1},
  pages = {013825},
  numpages = {10},
  year = {2009},
  month = {Jul},
  publisher = {American Physical Society},
  doi = {10.1103/PhysRevA.80.013825},
  url = {https://link.aps.org/doi/10.1103/PhysRevA.80.013825}
}

@article{Dorner2009,
  title = {Optimal Quantum Phase Estimation},
  author = {Dorner, U. and Demkowicz-Dobrzanski, R. and Smith, B. J. and Lundeen, J. S. and Wasilewski, W. and Banaszek, K. and Walmsley, I. A.},
  journal = {Phys. Rev. Lett.},
  volume = {102},
  issue = {4},
  pages = {040403},
  numpages = {4},
  year = {2009},
  month = {Jan},
  publisher = {American Physical Society},
  doi = {10.1103/PhysRevLett.102.040403},
  url = {https://link.aps.org/doi/10.1103/PhysRevLett.102.040403}
}

@Article{Zhou2018,
author={Zhou, Sisi
and Zhang, Mengzhen
and Preskill, John
and Jiang, Liang},
title={Achieving the Heisenberg limit in quantum metrology using quantum error correction},
journal={Nature Communications},
year={2018},
month={Jan},
day={08},
volume={9},
number={1},
pages={78},
issn={2041-1723},
doi={10.1038/s41467-017-02510-3},
url={https://doi.org/10.1038/s41467-017-02510-3}
}

@article{Fendley2004,
  title = {Competing density-wave orders in a one-dimensional hard-boson model},
  author = {Fendley, Paul and Sengupta, K. and Sachdev, Subir},
  journal = {Phys. Rev. B},
  volume = {69},
  issue = {7},
  pages = {075106},
  numpages = {15},
  year = {2004},
  month = {Feb},
  publisher = {American Physical Society},
  doi = {10.1103/PhysRevB.69.075106},
  url = {https://link.aps.org/doi/10.1103/PhysRevB.69.075106}
}

@article{Tóth_2014,
doi = {10.1088/1751-8113/47/42/424006},
url = {https://dx.doi.org/10.1088/1751-8113/47/42/424006},
year = {2014},
month = {oct},
publisher = {IOP Publishing},
volume = {47},
number = {42},
pages = {424006},
author = {Tóth, Géza and Apellaniz, Iagoba},
title = {Quantum metrology from a quantum information science perspective},
journal = {Journal of Physics A: Mathematical and Theoretical}
}

@article{Sidhu2020,
    author = {Sidhu, Jasminder S. and Kok, Pieter},
    title = {Geometric perspective on quantum parameter estimation},
    journal = {AVS Quantum Science},
    volume = {2},
    number = {1},
    pages = {014701},
    year = {2020},
    month = {02},
    issn = {2639-0213},
    doi = {10.1116/1.5119961},
    url = {https://doi.org/10.1116/1.5119961},
    eprint = {https://pubs.aip.org/avs/aqs/article-pdf/doi/10.1116/1.5119961/16700179/014701\_1\_online.pdf},
}

@article{Chai_2025,
doi = {10.1088/1572-9494/ada37c},
url = {https://dx.doi.org/10.1088/1572-9494/ada37c},
year = {2025},
month = {mar},
publisher = {IOP Publishing},
volume = {77},
number = {6},
pages = {065106},
author = {Chai, Qi and Yang, Wen},
title = {GHZ state, spin squeezed state, and spin coherent state for frequency estimation under general Gaussian noises},
journal = {Communications in Theoretical Physics}
}

@article{toth2012,
  title = {Multipartite entanglement and high-precision metrology},
  author = {T\'oth, G\'eza},
  journal = {Phys. Rev. A},
  volume = {85},
  issue = {2},
  pages = {022322},
  numpages = {8},
  year = {2012},
  month = {Feb},
  publisher = {American Physical Society},
  doi = {10.1103/PhysRevA.85.022322},
  url = {https://link.aps.org/doi/10.1103/PhysRevA.85.022322}
}

@article{Hyllus2012,
  title = {Fisher information and multiparticle entanglement},
  author = {Hyllus, Philipp and Laskowski, Wies\l{}aw and Krischek, Roland and Schwemmer, Christian and Wieczorek, Witlef and Weinfurter, Harald and Pezz\'e, Luca and Smerzi, Augusto},
  journal = {Phys. Rev. A},
  volume = {85},
  issue = {2},
  pages = {022321},
  numpages = {10},
  year = {2012},
  month = {Feb},
  publisher = {American Physical Society},
  doi = {10.1103/PhysRevA.85.022321},
  url = {https://link.aps.org/doi/10.1103/PhysRevA.85.022321}
}

@article{pezze2009,
  title = {Entanglement, Nonlinear Dynamics, and the Heisenberg Limit},
  author = {Pezz\'e, Luca and Smerzi, Augusto},
  journal = {Phys. Rev. Lett.},
  volume = {102},
  issue = {10},
  pages = {100401},
  numpages = {4},
  year = {2009},
  month = {Mar},
  publisher = {American Physical Society},
  doi = {10.1103/PhysRevLett.102.100401},
  url = {https://link.aps.org/doi/10.1103/PhysRevLett.102.100401}
}

@article{Jun2024,
  title = {Essay: Quantum Sensing with Atomic, Molecular, and Optical Platforms for Fundamental Physics},
  author = {Ye, Jun and Zoller, Peter},
  journal = {Phys. Rev. Lett.},
  volume = {132},
  issue = {19},
  pages = {190001},
  numpages = {11},
  year = {2024},
  month = {May},
  publisher = {American Physical Society},
  doi = {10.1103/PhysRevLett.132.190001},
}

@article{giovannetti2004,
    author = "Giovannetti, Vittorio and Lloyd, Seth and Maccone, Lorenzo",
    title = "{Quantum-Enhanced Measurements: Beating the Standard Quantum Limit}",
    doi = "10.1126/science.1104149",
    journal = "Science",
    volume = "306",
    number = "5700",
    pages = "1330--1336",
    year = "2004"
}

@article{Zanardi2006,
  title = {Ground state overlap and quantum phase transitions},
  author = {Zanardi, Paolo and Paunkovi\ifmmode \acute{c}\else \'{c}\fi{}, Nikola},
  journal = {Phys. Rev. E},
  volume = {74},
  issue = {3},
  pages = {031123},
  numpages = {6},
  year = {2006},
  month = {Sep},
  publisher = {American Physical Society},
  doi = {10.1103/PhysRevE.74.031123},
  url = {https://link.aps.org/doi/10.1103/PhysRevE.74.031123}
}

@Article{Aasi2013,
author={Aasi, J. and others},
title={Enhanced sensitivity of the LIGO gravitational wave detector by using squeezed states of light},
journal={Nat. Photonics},
year={2013},
month={Aug},
day={01},
volume={7},
number={8},
pages={613-619},
issn={1749-4893},
doi={10.1038/nphoton.2013.177},
}

@article{Tse2019,
  title = {Quantum-Enhanced Advanced LIGO Detectors in the Era of Gravitational-Wave Astronomy},
  author = {Tse, M. and others},
  journal = {Phys. Rev. Lett.},
  volume = {123},
  issue = {23},
  pages = {231107},
  numpages = {8},
  year = {2019},
  month = {Dec},
  publisher = {American Physical Society},
  doi = {10.1103/PhysRevLett.123.231107},
}

@article{Wasilewski2010,
  title = {Quantum Noise Limited and Entanglement-Assisted Magnetometry},
  author = {Wasilewski, W. and Jensen, K. and Krauter, H. and Renema, J. J. and Balabas, M. V. and Polzik, E. S.},
  journal = {Phys. Rev. Lett.},
  volume = {104},
  issue = {13},
  pages = {133601},
  numpages = {4},
  year = {2010},
  month = {Mar},
  publisher = {American Physical Society},
  doi = {10.1103/PhysRevLett.104.133601},
  url = {https://link.aps.org/doi/10.1103/PhysRevLett.104.133601}
}

@article{Wolfgramm2010,
  title = {Squeezed-Light Optical Magnetometry},
  author = {Wolfgramm, Florian and Cer\`e, Alessandro and Beduini, Federica A. and Predojevi\ifmmode \acute{c}\else \'{c}\fi{}, Ana and Koschorreck, Marco and Mitchell, Morgan W.},
  journal = {Phys. Rev. Lett.},
  volume = {105},
  issue = {5},
  pages = {053601},
  numpages = {4},
  year = {2010},
  month = {Jul},
  publisher = {American Physical Society},
  doi = {10.1103/PhysRevLett.105.053601},
  url = {https://link.aps.org/doi/10.1103/PhysRevLett.105.053601}
}

@article{Bollinger1996,
  title = {Optimal frequency measurements with maximally correlated states},
  author = {Bollinger, J. J . and Itano, Wayne M. and Wineland, D. J. and Heinzen, D. J.},
  journal = {Phys. Rev. A},
  volume = {54},
  issue = {6},
  pages = {R4649--R4652},
  numpages = {0},
  year = {1996},
  month = {Dec},
  publisher = {American Physical Society},
  doi = {10.1103/PhysRevA.54.R4649},
  url = {https://link.aps.org/doi/10.1103/PhysRevA.54.R4649}
}

@article{Monz2011,
  title = {14-Qubit Entanglement: Creation and Coherence},
  author = {Monz, Thomas and Schindler, Philipp and Barreiro, Julio T. and Chwalla, Michael and Nigg, Daniel and Coish, William A. and Harlander, Maximilian and H\"ansel, Wolfgang and Hennrich, Markus and Blatt, Rainer},
  journal = {Phys. Rev. Lett.},
  volume = {106},
  issue = {13},
  pages = {130506},
  numpages = {4},
  year = {2011},
  month = {Mar},
  publisher = {American Physical Society},
  doi = {10.1103/PhysRevLett.106.130506},
  url = {https://link.aps.org/doi/10.1103/PhysRevLett.106.130506}
}

@article{Tommaso2018,
  title = {Quantum Critical Metrology},
  author = {Fr\'erot, Ir\'en\'ee and Roscilde, Tommaso},
  journal = {Phys. Rev. Lett.},
  volume = {121},
  issue = {2},
  pages = {020402},
  numpages = {6},
  year = {2018},
  month = {Jul},
  publisher = {American Physical Society},
  doi = {10.1103/PhysRevLett.121.020402},
  url = {https://link.aps.org/doi/10.1103/PhysRevLett.121.020402}
}

@article{SisiZhou2025,
  doi = {10.22331/q-2025-06-05-1766},
  url = {https://doi.org/10.22331/q-2025-06-05-1766},
  title = {Stabilizer codes for {H}eisenberg-limited many-body {H}amiltonian estimation},
  author = {Antu, Santanu Bosu and Zhou, Sisi},
  journal = {{Quantum}},
  issn = {2521-327X},
  publisher = {{Verein zur F{\"{o}}rderung des Open Access Publizierens in den Quantenwissenschaften}},
  volume = {9},
  pages = {1766},
  month = jun,
  year = {2025}
}

@article{Bernien2017,
author={Bernien, Hannes
and Schwartz, Sylvain
and Keesling, Alexander
and Levine, Harry
and Omran, Ahmed
and Pichler, Hannes
and Choi, Soonwon
and Zibrov, Alexander S.
and Endres, Manuel
and Greiner, Markus
and Vuleti{\'{c}}, Vladan
and Lukin, Mikhail D.},
title={Probing many-body dynamics on a 51-atom quantum simulator},
journal={Nature},
year={2017},
month={Nov},
day={01},
volume={551},
number={7682},
pages={579-584},
issn={1476-4687},
doi={10.1038/nature24622},
url={https://doi.org/10.1038/nature24622}
}

@article{Ma2011,
  title = {Quantum Fisher information of the Greenberger-Horne-Zeilinger state in decoherence channels},
  author = {Ma, Jian and Huang, Yi-xiao and Wang, Xiaoguang and Sun, C. P.},
  journal = {Phys. Rev. A},
  volume = {84},
  issue = {2},
  pages = {022302},
  numpages = {7},
  year = {2011},
  month = {Aug},
  publisher = {American Physical Society},
  doi = {10.1103/PhysRevA.84.022302},
  url = {https://link.aps.org/doi/10.1103/PhysRevA.84.022302}
}

@article{Campos2007,
  title = {Quantum Critical Scaling of the Geometric Tensors},
  author = {Campos Venuti, Lorenzo and Zanardi, Paolo},
  journal = {Phys. Rev. Lett.},
  volume = {99},
  issue = {9},
  pages = {095701},
  numpages = {4},
  year = {2007},
  month = {Aug},
  publisher = {American Physical Society},
  doi = {10.1103/PhysRevLett.99.095701},
  url = {https://link.aps.org/doi/10.1103/PhysRevLett.99.095701}
}

@article{Schwandt2009,
  title = {Quantum Monte Carlo Simulations of Fidelity at Magnetic Quantum Phase Transitions},
  author = {Schwandt, David and Alet, Fabien and Capponi, Sylvain},
  journal = {Phys. Rev. Lett.},
  volume = {103},
  issue = {17},
  pages = {170501},
  numpages = {4},
  year = {2009},
  month = {Oct},
  publisher = {American Physical Society},
  doi = {10.1103/PhysRevLett.103.170501},
  url = {https://link.aps.org/doi/10.1103/PhysRevLett.103.170501}
}

@article{Albuquerque2010,
  title = {Quantum critical scaling of fidelity susceptibility},
  author = {Albuquerque, A. Fabricio and Alet, Fabien and Sire, Cl\'ement and Capponi, Sylvain},
  journal = {Phys. Rev. B},
  volume = {81},
  issue = {6},
  pages = {064418},
  numpages = {12},
  year = {2010},
  month = {Feb},
  publisher = {American Physical Society},
  doi = {10.1103/PhysRevB.81.064418},
  url = {https://link.aps.org/doi/10.1103/PhysRevB.81.064418}
}

@article{Montenegro2021,
  title = {Global Sensing and Its Impact for Quantum Many-Body Probes with Criticality},
  author = {Montenegro, Victor and Mishra, Utkarsh and Bayat, Abolfazl},
  journal = {Phys. Rev. Lett.},
  volume = {126},
  issue = {20},
  pages = {200501},
  numpages = {6},
  year = {2021},
  month = {May},
  publisher = {American Physical Society},
  doi = {10.1103/PhysRevLett.126.200501},
  url = {https://link.aps.org/doi/10.1103/PhysRevLett.126.200501}
}

@article{Mukhopadhyay2024,
  title = {Modular Many-Body Quantum Sensors},
  author = {Mukhopadhyay, Chiranjib and Bayat, Abolfazl},
  journal = {Phys. Rev. Lett.},
  volume = {133},
  issue = {12},
  pages = {120601},
  numpages = {7},
  year = {2024},
  month = {Sep},
  publisher = {American Physical Society},
  doi = {10.1103/PhysRevLett.133.120601},
  url = {https://link.aps.org/doi/10.1103/PhysRevLett.133.120601}
}

@article{Keesling_2019,
   title={Quantum Kibble–Zurek mechanism and critical dynamics on a programmable Rydberg simulator},
   volume={568},
   ISSN={1476-4687},
   url={http://dx.doi.org/10.1038/s41586-019-1070-1},
   DOI={10.1038/s41586-019-1070-1},
   number={7751},
   journal={Nature},
   publisher={Springer Science and Business Media LLC},
   author={Keesling, Alexander and Omran, Ahmed and Levine, Harry and Bernien, Hannes and Pichler, Hannes and Choi, Soonwon and Samajdar, Rhine and Schwartz, Sylvain and Silvi, Pietro and Sachdev, Subir and Zoller, Peter and Endres, Manuel and Greiner, Markus and Vuletić, Vladan and Lukin, Mikhail D.},
   year={2019},
   month=apr, pages={207–211} }

@article{Scholl_2023,
   title={Erasure conversion in a high-fidelity Rydberg quantum simulator},
   volume={622},
   ISSN={1476-4687},
   url={http://dx.doi.org/10.1038/s41586-023-06516-4},
   DOI={10.1038/s41586-023-06516-4},
   number={7982},
   journal={Nature},
   publisher={Springer Science and Business Media LLC},
   author={Scholl, Pascal and Shaw, Adam L. and Tsai, Richard Bing-Shiun and Finkelstein, Ran and Choi, Joonhee and Endres, Manuel},
   year={2023},
   month=oct, pages={273–278} }

@article{DiFresco2024,
  doi = {10.22331/q-2024-04-30-1326},
  url = {https://doi.org/10.22331/q-2024-04-30-1326},
  title = {Metrology and multipartite entanglement in measurement-induced phase transition},
  author = {Di Fresco, Giovanni and Spagnolo, Bernardo and Valenti, Davide and Carollo, Angelo},
  journal = {{Quantum}},
  issn = {2521-327X},
  publisher = {{Verein zur F{\"{o}}rderung des Open Access Publizierens in den Quantenwissenschaften}},
  volume = {8},
  pages = {1326},
  month = apr,
  year = {2024}
}

@Article{Giovanni2022,
	title={{Multiparameter quantum critical metrology}},
	author={Giovanni Di Fresco and Bernardo Spagnolo and Davide Valenti and Angelo Carollo},
	journal={SciPost Phys.},
	volume={13},
	pages={077},
	year={2022},
	publisher={SciPost},
	doi={10.21468/SciPostPhys.13.4.077},
	url={https://scipost.org/10.21468/SciPostPhys.13.4.077},
}

@misc{lee2022decoding,
      title={Decoding Measurement-Prepared Quantum Phases and Transitions: from Ising model to gauge theory, and beyond}, 
      author={Jong Yeon Lee and Wenjie Ji and Zhen Bi and Matthew P. A. Fisher},
      year={2022},
      eprint={2208.11699},
      archivePrefix={arXiv},
      primaryClass={cond-mat.str-el}
}

@article{Zhu2023Nov,
	author = {Zhu, Guo-Yi and Tantivasadakarn, Nathanan and Vishwanath, Ashvin and Trebst, Simon and Verresen, Ruben},
	title = {{Nishimori's Cat: Stable Long-Range Entanglement from Finite-Depth Unitaries and Weak Measurements}},
	journal = {Phys. Rev. Lett.},
	volume = {131},
	number = {20},
	pages = {200201},
	year = {2023},
	month = nov,
	publisher = {American Physical Society},
	doi = {10.1103/PhysRevLett.131.200201}
}

@article{Huang2024,
    author = {Huang, Jiahao and Zhuang, Min and Lee, Chaohong},
    title = {Entanglement-enhanced quantum metrology: From standard quantum limit to Heisenberg limit},
    journal = {Applied Physics Reviews},
    volume = {11},
    number = {3},
    pages = {031302},
    year = {2024},
    month = {07},
    issn = {1931-9401},
    doi = {10.1063/5.0204102},
    url = {https://doi.org/10.1063/5.0204102},
}

@article{Kitagawa1993,
  title = {Squeezed spin states},
  author = {Kitagawa, Masahiro and Ueda, Masahito},
  journal = {Phys. Rev. A},
  volume = {47},
  issue = {6},
  pages = {5138--5143},
  numpages = {0},
  year = {1993},
  month = {Jun},
  publisher = {American Physical Society},
  doi = {10.1103/PhysRevA.47.5138},
  url = {https://link.aps.org/doi/10.1103/PhysRevA.47.5138}
}

@article{Gietka2022,
  title = {Squeezing by critical speeding up: Applications in quantum metrology},
  author = {Gietka, Karol},
  journal = {Phys. Rev. A},
  volume = {105},
  issue = {4},
  pages = {042620},
  numpages = {12},
  year = {2022},
  month = {Apr},
  publisher = {American Physical Society},
  doi = {10.1103/PhysRevA.105.042620},
  url = {https://link.aps.org/doi/10.1103/PhysRevA.105.042620}
}

@misc{George2025,
      title={Critical Quantum Sensing: a tutorial on parameter estimation near quantum phase transitions}, 
      author={George Mihailescu and Uesli Alushi and Roberto Di Candia and Simone Felicetti and Karol Gietka},
      year={2025},
      eprint={2510.02035},
      archivePrefix={arXiv},
}

@article{Agarwal2025,
   title={Quantum sensing with ultracold simulators in lattice and ensemble systems: A review},
   note={Quantum sensing with ultracold simulators in lattice and ensemble systems: A review},
   ISSN={1793-6586},
   url={http://dx.doi.org/10.1142/S0129183125430065},
   journal={International Journal of Modern Physics C},
   publisher={World Scientific Pub Co Pte Ltd},
   author={Agarwal, Keshav Das and Mondal, Sayan and Sahoo, Ayan and Rakshit, Debraj and Sen(De), Aditi and Sen, Ujjwal},
   year={2025},
   month=dec }

@article{sarkar2025,
      author={Anindita Sarkar and Paranjoy Chaki and Priya Ghosh and Ujjwal Sen},
      title={Fluctuation in energy extraction from quantum batteries: How open should the system be to control it?
},
      year={2025},
      eprint={2510.24484},
      archivePrefix={arXiv},
      primaryClass={quant-ph},
      url={https://arxiv.org/abs/2510.24484},
}

@inbook{De_Pasquale_2018,
   title={Quantum Thermometry},
   ISBN={9783319990460},
   ISSN={2365-6425},
   url={http://dx.doi.org/10.1007/978-3-319-99046-0_21},
   DOI={10.1007/978-3-319-99046-0_21},
   booktitle={Thermodynamics in the Quantum Regime},
   publisher={Springer International Publishing},
   author={De Pasquale, Antonella and Stace, Thomas M.},
   year={2018},
   pages={503–527} }

@article{Rubio2021,
  title = {Global Quantum Thermometry},
  author = {Rubio, Jes\'us and Anders, Janet and Correa, Luis A.},
  journal = {Phys. Rev. Lett.},
  volume = {127},
  issue = {19},
  pages = {190402},
  numpages = {6},
  year = {2021},
  month = {Nov},
  publisher = {American Physical Society},
  doi = {10.1103/PhysRevLett.127.190402},
  url = {https://link.aps.org/doi/10.1103/PhysRevLett.127.190402}
}

@article{Kurdzia2023,
  title = {Using Adaptiveness and Causal Superpositions Against Noise in Quantum Metrology},
  author = {Kurdzia\l{}ek, Stanis\l{}aw and G\'orecki, Wojciech and Albarelli, Francesco and Demkowicz-Dobrza\ifmmode \acute{n}\else \'{n}\fi{}ski, Rafa\l{}},
  journal = {Phys. Rev. Lett.},
  volume = {131},
  issue = {9},
  pages = {090801},
  numpages = {7},
  year = {2023},
  month = {Aug},
  publisher = {American Physical Society},
  doi = {10.1103/PhysRevLett.131.090801},
  url = {https://link.aps.org/doi/10.1103/PhysRevLett.131.090801}
}

@article{Kurdzia2025,
  title = {Universal Bounds for Quantum Metrology in the Presence of Correlated Noise},
  author = {Kurdzia\l{}ek, Stanis\l{}aw and Albarelli, Francesco and Demkowicz-Dobrza\ifmmode \acute{n}\else \'{n}\fi{}ski, Rafa\l{}},
  journal = {Phys. Rev. Lett.},
  volume = {135},
  issue = {13},
  pages = {130801},
  numpages = {7},
  year = {2025},
  month = {Sep},
  publisher = {American Physical Society},
  doi = {10.1103/jy3v-wkcb},
  url = {https://link.aps.org/doi/10.1103/jy3v-wkcb}
}

@article{yinan2025,
  title = {Bosonic entanglement and quantum sensing from energy transfer in two-tone Floquet systems},
  author = {Chen, Yinan and Elben, Andreas and Rubio, Angel and Refael, Gil},
  journal = {Phys. Rev. Res.},
  volume = {7},
  issue = {4},
  pages = {043014},
  numpages = {36},
  year = {2025},
  month = {Oct},
  publisher = {American Physical Society},
  doi = {10.1103/cm49-smhr},
  url = {https://link.aps.org/doi/10.1103/cm49-smhr}
}

@article{williamson_low-overhead_2026,
	title = {Low-overhead fault-tolerant quantum computation by gauging logical operators},
	volume = {22},
	issn = {1745-2481},
	url = {https://doi.org/10.1038/s41567-026-03220-8},
	doi = {10.1038/s41567-026-03220-8},
	number = {4},
	journal = {Nature Physics},
	author = {Williamson, Dominic J. and Yoder, Theodore J.},
	month = apr,
	year = {2026},
	pages = {598--603},
}

@misc{chen2026quantum,
      title={Quantum metrology via partial quantum error correction}, 
      author={Yinan Chen and Zongyuan Wang and Sisi Zhou},
      year={2026},
      eprint={2605.08341},
      archivePrefix={arXiv},
      primaryClass={quant-ph},
      url={https://arxiv.org/abs/2605.08341}, 
}

@article{W2014,
  title = {Improved Quantum Metrology Using Quantum Error Correction},
  author = {D\"ur, W. and Skotiniotis, M. and Fr\"owis, F. and Kraus, B.},
  journal = {Phys. Rev. Lett.},
  volume = {112},
  issue = {8},
  pages = {080801},
  numpages = {5},
  year = {2014},
  month = {Feb},
  publisher = {American Physical Society},
  doi = {10.1103/PhysRevLett.112.080801}
}

@article{
Zou2018,
author = {Yi-Quan Zou  and Ling-Na Wu  and Qi Liu  and Xin-Yu Luo  and Shuai-Feng Guo  and Jia-Hao Cao  and Meng Khoon Tey  and Li You },
title = {Beating the classical precision limit with spin-1 Dicke states of more than 10,000 atoms},
journal = {Proceedings of the National Academy of Sciences},
volume = {115},
number = {25},
pages = {6381-6385},
year = {2018},
doi = {10.1073/pnas.1715105115},
URL = {https://www.pnas.org/doi/abs/10.1073/pnas.1715105115},
eprint = {https://www.pnas.org/doi/pdf/10.1073/pnas.1715105115}}

@misc{musso2026,
      title={Quantum Fisher Information under decoherence with explicit wavefunctions}, 
      author={Francesco Musso and Vittorio Vitale and Sara Murciano},
      year={2026},
      eprint={2605.22917},
      archivePrefix={arXiv},
      primaryClass={quant-ph},
      url={https://arxiv.org/abs/2605.22917}, 
}

 \newpage
\appendix
\section{Connetion with Ref.~\cite{frerot2024symmetry}}\label{app:connection}

In this appendix we clarify the relation between the results obtained in Ref.~\cite{frerot2024symmetry} and the symmetry-informed interferometric quantum-sensing protocol discussed in the main text. The point is that the latter can be viewed as an equivalent formulation in terms of a $\mathbb{Z}_2$ parity, after removing a part of the imprinting operator that does not affect the metrological response of states supported in the chosen sector.
Let $P$ be a projector onto the subspace containing the probe state $\rho$, so that $P\rho=\rho P=\rho$.
Reference~\cite{frerot2024symmetry} considers an imprinting generator $O$ satisfying $POP=0$.
This condition means that the action of $O$ on states in the $P$ sector takes them entirely outside that sector. Introducing the complementary projector $Q=1-P$, we can decompose the operator $O$ into blocks with respect to the decomposition of the Hilbert space $\mathcal{H}=P\mathcal{H}\oplus Q\mathcal{H}$. Since $POP=0$, one has $O=POQ+QOP+QOQ$.
The term $QOQ$ acts only within the orthogonal complement of the support of $\rho$. It therefore does not affect the second-order response of the state $\rho$ inside the $P$ sector. Thus, we introduce the chiralized generator
\begin{equation}
    O_{\rm ch}=POQ+QOP .
\end{equation}
We now define the operator $\mathcal{A}=2P-1=P-Q$ satisfying $\mathcal{A}^2=1$.
It defines a canonical $\mathbb{Z}_2$ decomposition of the Hilbert space: states in the $P$ sector have $\mathcal{A}=+1$, while states in the $Q$ sector have $\mathcal{A}=-1$. Since the probe state is supported in the $P$ sector, $\mathcal{A}\rho=\rho \mathcal{A}=\rho$.
Moreover, the chiralized operator is purely off-diagonal with respect to the decomposition $P\oplus Q$, i.e., one can easily prove that $\{\mathcal{A},O_{\rm ch}\}=0$.
\\ It remains to check that the replacement of $O$ by $O_{\rm ch}$ does not modify the metrologically relevant second moment. Since $POP=0$, we can prove that $PO^2P=PO^2_{\rm ch}P$.
\\Thus, the result of Ref.~\cite{frerot2024symmetry} gives a group-independent projector criterion:
if the probe state is supported in a symmetry sector $P$ and the imprinter
has no matrix elements within that sector, $POP=0$, then measuring the
sector projector saturates the local metrological response. Our result can be viewed as a practical way of
ensuring this condition from symmetry. In the $\mathbb Z_2$ case, the
condition $\{A,O\}=0$ implies that $O$ flips the symmetry sector of the
probe state, and therefore $POP=0$. For a general finite group, the analogous selection rule is that the \textit{metrologically relevant} component of $O$ carries a nontrivial symmetry charge. As a result, it maps the sector supporting the probe state only to orthogonal sectors. For a general imprinter, the chiralized operator $O_{\rm ch}$ precisely filters for the sector-changing component that preserves the local second-order metrological response, while making the symmetry structure explicit.

\section{Mean square error and Fisher information}
\label{mean square error}

In this appendix, we prove that for the parity measurement $\mathcal{A}=\prod_{j}X_{j}$, the mean square error in $\braket{\mathcal{A}}$ equals the classical Fisher information of the corresponding projective measurement, described by $\mathcal{P}_{\pm}=\frac{1}{2}\left(\mathbb{I}\pm\mathcal{A}\right)$. Accordingly, the probability of an outcome $\pm$ is given as $P_{\pm}=\braket{\mathcal{P}_{\pm}}=\frac{1}{2}\left(1\pm\braket{\mathcal{A}}\right)$. Inserting $P_{\pm}$ into the defining equation for the classical Fisher information, we obtain
\begin{align}
    F_{c}=\frac{1}{P_{+}}\left(\partial_{\theta}P_{+}\right)^2+\frac{1}{P_{-}}\left(\partial_{\theta}P_{-}\right)^2 =\frac{1}{2}\left[\frac{1}{1+\braket{\mathcal{A}}}+\frac{1}{1-\braket{\mathcal{A}}}\right]\left(\partial_{\theta}\mathcal{A}\right)^2=\frac{\left(\partial_{\theta}\mathcal{A}\right)^2}{1-\braket{\mathcal{A}}^2}=\delta\theta^{-2}.
\end{align}
\noindent In the last line, we use $\mathcal{A}^2=\mathbb{I}$ for parity. The proof hence directly generalizes to other $\mathbb{Z}_{2}$ symmetries.

\section{Global and local dephasing for a critical state}
\label{Global and local dephasing}

In this appendix, we prove that the precision of the spin measurement in the $xy-$plane under global and local dephasing follows the SQL scaling in the system size $L$. 

The global dephasing noise corresponds to measure a global noisy longitudinal magnetic field
\begin{equation}
    B(t)=B+\Tilde{B}(t),
\end{equation}
where $B$ is the time-independent mean value of the field and $\Tilde{B}(t)$ is the time-dependent Gaussian fluctuation characterized by
\begin{equation}
    C(t)=\overline{\Tilde{B}(t)\Tilde{B}(0)},
\end{equation}
where $\overline{\square}$ is the ensemble average of the fluctuation. The critical state is then mapped to \cite{Chai_2025}
\begin{equation}
    \rho_{c}=e^{-iBt\sum_{j}Z_{j}}K_{t}\left\{\ket{\psi}_{c}\bra{\psi}_{c}\right\}e^{iBt\sum_{j}Z_{j}}
\end{equation}
with $K_{t}\left\{\rho\right\}=\overline{e^{-i\Tilde{\phi}(t)\sum_{j}Z_{j}}\rho e^{i\Tilde{\phi}(t)\sum_{j}Z_{j}}}$ and $\Tilde{\phi}(t)=\int\limits_{0}^{t}d\tau\Tilde{B}(\tau)$. To learn about $B$, we apply a measurement on the total spin in the xy-plane (at the azimuthal angle $\theta$) $S_{\theta}=\frac{1}{2}\sum_{j}X_{j}\cos{\theta}+\frac{1}{2}\sum_{j}Y_{j}\sin{\theta}$, whose sensitivity $\delta B$ is given by the Cramer-Rao relation
\begin{equation}
    \delta B=\frac{\delta S_{\theta}}{\left|\partial_{B}\braket{S_{\theta}}_{c}\right|},\label{B-sensitivity}
\end{equation}
where $\braket{\square}_{c}=\mathrm{Tr}\left\{\rho_{c}\square\right\}$ and $\delta S_{\theta}=\sqrt{\braket{S_{\theta}^{2}}_{c}-\braket{S_{\theta}}_{c}^{2}}$. By applying the conjugate channel $K^{\dagger}\left\{\square\right\}=\overline{e^{i\Tilde{\phi}(t)\sum_{j}Z_{j}}\square e^{-i\Tilde{\phi}(t)\sum_{j}Z_{j}}}$ to $S_{\theta}$, the mean and variance of $S_{\theta}$ can be equivalently evaluated as
\begin{align*}
    \braket{S_{\theta}}_{c}&=\bra{\psi}_{c}e^{iBt\sum_{j}Z_{j}}K_{t}^{\dagger}\left\{S_{\theta}\right\}e^{-iBt\sum_{j}Z_{j}}\ket{\psi}_{c},\\
    \braket{S_{\theta}^{2}}_{c}&=\bra{\psi}_{c}e^{iBt\sum_{j}Z_{j}}K_{t}^{\dagger}\left\{S_{\theta}^{2}\right\}e^{-iBt\sum_{j}Z_{j}}\ket{\psi}_{c}.
\end{align*}
Noting $\frac{1}{2}\sum_{j}Z_{j}$ commutes with $\left(\frac{1}{2}\sum_{j}X_{j}\right)^2+\left(\frac{1}{2}\sum_{j}Y_{j}\right)^2$, the action of channel $K_{t}^{\dagger}$ can be explicitly given as:
\begin{align*}
    K_{t}^{\dagger}\left\{S_{\theta}\right\}&=e^{-\chi(t)}S_{\theta},\\
    K_{t}^{\dagger}\left\{S_{\theta}^{2}\right\}&=e^{-4\chi(t)}S_{\theta}^{2}+\frac{1}{2}\left(1-e^{-4\chi(t)}\right)\left(S_{x}^{2}+S_{y}^{2}\right),
\end{align*}
where we have introduced $S_{x}=\frac{1}{2}\sum_{j}X_{j}$, $S_{y}=\frac{1}{2}\sum_{j}Y_{j}$, and $\chi(t)=4\int_{0}^{t}d\tau\left(t-\tau\right)C(\tau)$. The sensitivity $\delta B$ follows as
\begin{equation}
    \frac{\sqrt{e^{-4\chi(t)}\braket{S_{\theta-2Bt}^{2}}+\frac{1}{2}\left(1-e^{-4\chi(t)}\right)\braket{S_{x}^{2}+S_{y}^{2}}-e^{-2\chi(t)}\braket{S_{\theta-2Bt}}^{2}}}{2e^{-\chi(t)}\left|\partial_{B}\braket{S_{\theta-2Bt}}\right|},
\end{equation}
\noindent where $\braket{\square}=\bra{\psi}_{c}\square\ket{\psi}_{c}$. For the critical Ising model, $\braket{S_{x}}=\frac{1}{\pi}L$, $\braket{S_{y}}=\braket{S_{z}}=0$, $\braket{S_{x}^{2}}=C_{x}L$, and $\braket{S_{y}^{2}}=C_{y}L$. We then have the best sensitivity occuring at $\theta=2Bt+\frac{\pi}{2}$ as
\begin{equation}
    \delta B=\frac{\pi}{2t\sqrt{L}}\sqrt{e^{-2\chi(t)}C_{y}+\frac{1}{2}\left[e^{2\chi(t)}-e^{-2\chi(t)}\right]\left(C_{x}+C_{y}\right)},\label{spin sensitivity of critical}
\end{equation}
which, comparing to the sensitivity of the GHZ state $\delta B=\frac{e^{L\chi(t)}}{2Lt}$, shows a slower increase in $t$ and the SQL scaling in $N$. For local dephasing, a similar proof is already present in Section \ref{sec:dephasing} and we retrieve the same result here.

We also note that the sensitivity of the critical state Eq.~\eqref{spin sensitivity of critical} is as well better than that of the spin-coherent state, while becomes worse than that of the spin-squeezed state.

We next consider the sensitivity of critical states under local dephasing, which can be similarly described by 
\begin{equation}
    \Tilde{K}_{t}\left\{\rho\right\}=\prod_{j}k_{j,t}\left\{\rho\right\},
\end{equation}
with $k_{j,t}\left\{\rho\right\}=\overline{e^{-i\Tilde{\phi}(t)Z_{j}}\rho e^{i\Tilde{\phi}(t)Z_{j}}}$. As before, we define the conjugate channel as $k_{j,t}^{\dagger}\left\{\rho\right\}=\overline{e^{i\Tilde{\phi}(t)Z_{j}}\rho e^{-i\Tilde{\phi}(t)Z_{j}}}$. Noting that $k_{j,t}^{\dagger}\left\{X_{j}\right\}=e^{-\chi(t)}X_{j}$ and $k_{j,t}^{\dagger}\left\{Y_{j}\right\}=e^{-\chi(t)}Y_{j}$ \footnote{We note that the action of $k^{\dagger}_{j,t}$ is equivalent to the local dephasing channel $\mathcal{E}_{j}[\rho]=(1-p)\mathbb{I}+pZ_{j}\rho Z_{j}$ by setting $p=\frac{1-e^{-\chi(t)}}{2}$}, we obtain
\begin{align*}
    \Tilde{K}_{t}^{\dagger}\left\{S_{\theta}\right\}&=e^{-\chi(t)}S_{\theta},\\
    \Tilde{K}_{t}^{\dagger}\left\{S_{\theta}^{2}\right\}&=e^{-2\chi(t)}S_{\theta}^{2}+\frac{L}{4}\left(1-e^{-2\chi(t)}\right),   
\end{align*}
where $\Tilde{K}_{t}^{\dagger}=\prod_{j}k_{j,t}^{\dagger}$ is the conjugate channel of $\Tilde{K}_{t}$. The sensitivity of the spin measurement follows as
\begin{equation}
    \delta B=\frac{\sqrt{e^{-2\chi(t)}\braket{S_{\theta-2Bt}^{2}}+\frac{L}{4}\left[1-e^{-2\chi(t)}\right]-e^{-2\chi(t)}\braket{S_{\theta-2Bt}}^{2}}}{2e^{-\chi(t)}\left|\partial_{B}\braket{S_{\theta-2Bt}}\right|}.
\end{equation}
For the critical Ising model, again, we reach the maximal sensitivity at $\theta=2Bt+\frac{\pi}{2}$ as
\begin{equation}
    \delta B=\frac{\pi}{2t\sqrt{L}}\sqrt{C_{y}+\frac{1}{4}\left[e^{\chi(t)}-1\right]},
\end{equation}
which exhibits a SQL behavior in $L$. We also note the sensitivity of the GHZ state drops exponentially as $\delta B=\frac{e^{L\chi(t)}}{2Lt}$. Consequently, the critical states as well perform better than the GHZ state, sharing a similar SQL behavior as the spin-coherent state. 

Finally, we remark the equivalence of this noisy evolution and the Kraus method for the local dephasing described in Section \ref{sec:dephasing} by setting $p(t)=\frac{1-e^{-\chi(t)}}{2}$, and the dephasing on the $j-$th qubit is given by $\mathcal{E}_{j}[\rho]=[1-p(t)]\rho+p(t)Z_{j}\rho Z_{j}$. More explicitly, from the error propagation formula, the relevant quantities of Section \ref{sec:dephasing} are $\braket{S_{\theta}}=\mathrm{Tr}\left\{\prod_{j}\mathcal{E}_{j}[\rho_{0}] S_{\theta}\right\}=\mathrm{Tr}\left\{\rho_{0}\prod_{j}\mathcal{E}^{*}_{j}[S_{\theta}]\right\}$ and $\braket{S_{\theta}^{2}}=\mathrm{Tr}\left\{\prod_{j}\mathcal{E}_{j}[\rho_{0}] S_{\theta}^{2}\right\}=\mathrm{Tr}\left\{\rho_{0}\prod_{j}\mathcal{E}^{*}_{j}[S_{\theta}^{2}]\right\}$. Here we have used the fact that, since correlators are linear objects in $\rho$, the action of $\mathcal{E}[\cdot]=\prod_{j}\mathcal{E}_{j}[\cdot]$ on $\rho$ can be equivalently computed as the action of its conjugate channel $\mathcal{E}^*[\cdot]$ (which in this case agrees with $\mathcal{E}[\cdot]$) on $\sum_{j}X_j$ or $\sum_{j}Y_j$.
Using $\mathcal{E}_{j}^{*}\left[X_{k}\right]=(1-2p\ \delta_{j,k})X_{k}$ and $\mathcal{E}_{j}^{*}\left[Y_{k}\right]=(1-2p\ \delta_{j,k})Y_{k}$, we find

\begin{align}
    \prod_{j}\mathcal{E}^{*}_{j}\left[S_{\theta}\right]&=(1-2p)S_{\theta},\\
    \prod_{j}\mathcal{E}^{*}_{j}\left[S_{\theta}^{2}\right]&=(1-2p)^{2}S_{\theta}^{2}+p(1-p)L.
\end{align}

\section{Bounds on the QFI for decohered critical states}

In this appendix, we review different bounds on the QFI found in recent literature and evaluate their performance when applied to decohered critical states.

We begin with the monotonic sequence $F_{n}$ introduced in \cite{Rath2021}:
\begin{equation}
    F_{n}=2\sum_{\alpha, \beta}\sum_{l=0}^{n}\left(\lambda_{i}-\lambda_{j}\right)^2(1-\lambda_{i}-\lambda_{j})^{l}\left|\bra{i}O\ket{j}\right|^{2},
\end{equation}
where $\lambda_{i}$, $\ket{i}$ are the $i$-th eigenvalue and the corresponding eigenstate of the density matrix $\rho$. The sequence of $F_n$ satisfy the inequality
\begin{equation}
    F_{n}\leq F_{n+1}\leq 2F_{n} \label{eq: Fn_sandwich}.
\end{equation}
The first inequality follows from $\lambda_{i}+\lambda_{j}\leq1$, which also guarantees that
\begin{equation}
    (1-\lambda_{i}-\lambda_{j})^{n+1}\leq(1-\lambda_{i}-\lambda_{j})^{n}.
\end{equation}
This relation constrains the convergence rate of the series 
$F_n$ toward the QFI as $n\rightarrow\infty$. Typically, if $F_{n}$ scales with system size as $F_{n}\sim L^{\Delta}$, the same scaling holds for any finite $m$. Otherwise, if $F_{n}$ decays exponentially, then any other $F_{m}$ with finite $m$  will also decays exponentially, as it follows from the inequality \eqref{eq: Fn_sandwich}

For decohered critical states, we generally observe the latter scenario, i.e., $F_n$ decays exponentially with system size. This is already evident at $n=0$ where $F_{0}=4\mathrm{Tr}\left\{\rho\left[\rho,O\right]O\right\}$. At this point, we can introduce a more useful quantity
\begin{equation}
D_{2}=4\frac{\mathrm{Tr}\left\{\rho\left[\rho,O\right]O\right\}}{\mathrm{Tr}\left\{\rho^2\right\}},
\end{equation}
and later in the appendix we will show that $D_{2}\sim L^{\Delta}$ does exhibit polynomial scaling with $L$. Thus, 
\begin{equation}
F_{0}\sim \mathrm{Tr}\left\{\rho^2\right\}D_{2}\sim\mathrm{Tr}\left\{\rho^2\right\}L^{\Delta}.
\end{equation}
However, for a uniformly decohered critical state, the purity $\mathrm{Tr}\left\{\rho^2\right\}$ decays exponentially. This implies that $F_0$ itself decays exponentially with 
$L$, and by inequality~\eqref{eq: Fn_sandwich}, so does every finite $F_{n}$.

We now explore the scaling of $D_{2}$. As proposed in \cite{Xu2023}, $D_2$ can serve as an order parameter for the phase transition of the decohered critical system. It also appears as the second term in the $n$-th Jeffreys distance
\begin{align}
    D^{(n)}(\rho,\sigma)\equiv&\frac{1}{n-1}\bigg[\log \mathrm{Tr}(\rho^n)+\log\mathrm{Tr}(\sigma^{n})\\
    &-\log\mathrm{Tr}(\rho\sigma^{n-1})-\log\mathrm{Tr}(\sigma\rho^{n-1})\bigg].
\end{align}
By setting $\sigma\rightarrow e^{iO\theta}\rho e^{-iO\theta}$, one finds $D_{2}=4\partial_{\theta}^{2}D^{(2)}$. Crucially, in the limit $n\rightarrow1$, $D^{(n)}$ converges to the quantum generalization of the Jeffreys divergence and $\partial_{\theta}^{2}D^{(n\rightarrow1)}$ recovers the classical Fisher information.

Although $D_2$ is sometimes interpreted as a generalized metric for metrological performance, its scaling with $L$ may differ from that of the QFI. We demonstrate this below.

Consider the ground state of the critical Ising model, and define the operator  $O=\sum_j Z_j$. After applying uniformly a specific quantum channel on each qubit $\mathcal{E}(\cdot)=\prod_{j}\mathcal{E}_{j}(\cdot)$, the pristine critical state is mapped to the mixed state in Eq.~\eqref{eq:ZZ}.
To proceed further, we apply the Choi-Jamiolkowski isomorphism to map the density matrix $\rho_0=\ket{\psi}\bra{\psi}$ into the pure state $\ket{\rho_0}\rangle=\ket{\psi}\ket{\psi^*}$ and an arbitrary positive channel $\mathcal{E}=\prod_j\mathcal{E}_j$ into the operator in the doubled Hilbert space. The Choi operator of the channel is given by \cite{Yuto24}
\begin{equation}
    \hat{\mathcal{E}}\propto e^{-\mu \sum_j \hat{k}_j\otimes \hat{\tilde{k}}_j},
\end{equation}
where $k_j$ and $\hat{\tilde{k}}_j$ are local operators depending on the specific quantum channel we are applying and $\mu=\mathrm{arctanh} [p/(1-p)]$ measures the amount of dissipation or decoherence the system is subject to. 
Therefore, by exploiting the exponential form of $\hat{\mathcal{E}}$, the vectorized operator becomes $\ket{\rho}\rangle=\hat{\mathcal{E}}\ket{\rho_0}\rangle    $ (properly normalized). In the continuum formulation of the problem, we can write down the matrix elements of the doubled density matrix $\hat{\rho}^D=\ket{\rho}\rangle\langle \bra{\rho}$ as \cite{Yuto24}
\begin{align}
\bra{\phi'(x),\tilde{\phi}'(x)}  \hat{\rho}^D\ket{\phi''(x),\tilde{\phi}''(x)}=
\frac{1}{Z}\displaystyle \int_{(\phi,\tilde{\phi})_{\tau=0^+}=(\phi'',\tilde{\phi}'')}^{(\phi,\tilde{\phi})_{\tau=0^-}=(\phi',\tilde{\phi}')} \mathcal{D}\phi \mathcal{D}\tilde{\phi}e^{-\mathcal{S}^{\mathcal{E}}[\phi,\tilde{\phi}]}, 
\end{align}
where $\phi$ and $\tilde{\phi}$ live in the doubled Hilbert space, and the total action reads 
\begin{equation}
\mathcal{S}^{\mathcal{E}}[\phi,\tilde{\phi}]=\mathcal{S}_0[\phi]+\mathcal{S}_0[\tilde{\phi}]+\mathcal{S}_{\mathcal{E}}[\phi,\tilde{\phi}].
\end{equation}
Here $\mathcal{S}_0$ is the bulk action of the ground state $\ket{\psi_0}$, $\mathcal{S}_{\mathcal{E}}$
represents the effect of the channel, and it can be written as a boundary term acting on the $\tau = 0$ line. 

In this vectorized formalism, the quantity we are interested in reads
\begin{align}\label{eq:field}
&\frac{\mathrm{Tr}[\rho^2O^2]-\mathrm{Tr}[\rho\, O \rho\, O]}{\mathrm{Tr}[\rho^2]}=\langle 
\braket{\rho|O^2\otimes \mathbbm{1}|\rho}\rangle-\langle 
\braket{\rho|O\otimes \tilde{O}|\rho}\rangle\\
&=2\sum_{j<k}\langle 
\braket{\rho|Z_j Z_k\otimes \mathbbm{1}|\rho}\rangle-\sum_{j\neq k}\langle 
\braket{\rho|Z_j\otimes \tilde{Z}_k|\rho}\rangle.
\end{align}
Let us focus on $\ket{\psi_0}$ being the ground state of the critical Ising model, and $\mathcal{E}_j=(1-p)\rho_0+pZ_j\rho_0Z_j$. In this case, the defect line is 
\begin{equation}
    \mathcal{S}_{\mathcal{E}}=\mu \int_x \sigma_{\tau=0} \tilde{\sigma}_{\tau=0}.
\end{equation}
This term is a relevant perturbation, so it will drive the system towards the Ising boundary criticality. This means that each term in Eq. 
\eqref{eq:field} scales at most as $O(L)$, even though with different prefactors due to the different regularization of each term. Indeed, in $\langle 
\braket{\rho|Z_j Z_k\otimes \mathbbm{1}|\rho}\rangle$ we have to consider a lattice cutoff within the same copy of the Hilbert space, while in the term $\langle 
\braket{\rho|Z_j \otimes Z_k|\rho}\rangle$ it will involve different copies.

If the channel contains $\mathcal{E}_j=(1-p)\rho_0+p X_j \rho_0 X_j$, the action would be 
\begin{equation}\label{eq:marginal}   \mathcal{S}_{\mathcal{E}}=b(\mu \braket{X})\int_x (\varepsilon+\tilde{\varepsilon})_{\tau=0},
\end{equation}
where $b(\mu, \braket{X})$ is a function depending on the measurement strength and a non-universal constant due to the continuum limit we are considering. In our case, $b(\mu, \braket{X})\propto \mathrm{arctanh}(p/(1-p))$. The form of the interaction term \eqref{eq:marginal} implies that the 2 copies of the Hilbert space remain decoupled, so the second term in Eq.~\eqref{eq:field} simply vanishes and the first term would scale as 
\begin{equation}
    L^{2(1-\Delta_p)}, \quad \Delta_{p}=\frac{2\arctan^2(e^{8/\pi \mathrm{arctanh}(p/(1-p))})}{\pi^2}. 
\end{equation}
The explicit form of $\Delta_{p}$ can be derived from the scaling behavior of the order parameter in one single copy of the Ising model in the presence of the marginal defect line $\varepsilon$ [].
Since $\Delta_{p}\in [1/8,1/2]$, this result implies a clear $p$ dependence of the critical exponent $\Delta_{p}$, whereas in the main text we prove that the QFI always scales as $L^{2(1-1/8)}$ for $p<0.5$.

A third case that we can examine is the one where $\mathcal{E}_j=(1-p)\rho_0+p Z_jZ_{j+1}\rho_0 Z_jZ_{j+1}$. Since $Z_jZ_{j+1}$ also maps to $-\varepsilon$, the defect line becomes 
\begin{equation} \mathcal{S}_{\mathcal{E}}=-b(\mu \braket{X})\int_x (\varepsilon+\tilde{\varepsilon})_{\tau=0},
\end{equation}
which is the same as Eq.~\eqref{eq:marginal} up to an overall sign. Therefore, also in this case the second contribution in Eq.~\eqref{eq:field} vanishes and the first term would scale as $L^{2(1-\Delta_{p})}$ where now $\Delta_{p}=2/\pi^2\arctan^2(e^{-8/\pi \mathrm{arctanh}(p/(1-p))})\in [0,1/8]$.
\section{Analytical details about local spin flip }\label{app:spinflip}

In this appendix, we derive the analytical expression for the QFI in the presence of local spin flip. The action of the uniform qubit flip channel on the critical state yields $\rho=\sum_{E}p^{\left|E\right|}(1-p)^{L-\left|E\right|}X_{E}\rho_{0} X_{E}$, with $E$ encoding a specific error configuration where the channel acts non-trivially, such that $X_{E}=\prod_{i\in E}X_i$ is the product of bit flips on all sites belonging to $E$, and $|E|$ corresponding to the size of this set.
To compute $\delta\theta^{-2}$ [cf. Eq.~\eqref{eq:meansquare}], we expand both the denominator and the numerator in $\theta$ and take the limit $\theta\rightarrow0$. Crucially, as $[X_{E},\prod_{i}X_{i}]=0$ for the parity operator, a parallel calculation from Eq.~\eqref{eq:var_theta} to Eq.~\eqref{eq:main1} yields
\begin{equation}\label{eq:resolution_X}
\delta\theta^{-2}=4\sum_{E}p^{\left|E\right|}(1-p)^{L-\left|E\right|}\braket{X_{E}O^2 X_{E}}_0,
\end{equation}
with $O=\sum_{i}Z_{i}$ and $\braket{\cdot}_0=\mathrm{Tr}\left\{\rho_{0}\ \cdot\right\}$. From the CR bound, the result above represents a lower bound to the QFI since $F_{Q}[\rho]\geq\delta\theta^{-2}$.

At the same time, we can also show that $\delta\theta^{-2}$ is an upper bound on the QFI. From the convexity of the QFI, we have $F_{Q}[\sum_{i}\rho_{i}]\leq\sum_{i}F_{Q}[\rho_{i}]$. An upper bound of $F_{Q}[\rho]$ is then given as:
\begin{equation}\label{eq:upperbound}
    F_{Q}[\rho]\leq\sum_{E}p^{\left|E\right|}(1-p)^{L-\left|E\right|} F_{Q}[X_{E}\rho_{0}X_{E}].
\end{equation}
Since the state $X_{E}\rho_{0}X_{E}$ is pure, its QFI equals $4$ times the variance $F_{Q}[X_{E}\rho_{0}X_{E}]=4\braket{X_{E}O^2 X_{E}}_0$, according to Eq.~\eqref{eq:QFIpure}. Plugging the result in Eq.~\eqref{eq:upperbound}, we find $F_{Q}[\rho]\leq\delta\theta^{-2}$. Therefore, we can conclude that the parity measurement is optimal also under the presence of decoherence due to bit flips, which can be modeled by the $X$ quantum channel. 

We can exploit the fact that the parity is the optimal measurement to compute the QFI under different dissipation strength, $p$. Due to its optimality, a parallel derivation from Eq.~\eqref{eq:meansquare} to Eq.~\eqref{eq:main1} yields $F_{Q}[\rho]=4 \mathrm{Tr}(O^2\mathcal{E}[\rho])$. As the variance is a correlation function which is a linear object in $\rho$, the action of $\mathcal{E}[\cdot]=\prod_{j}\mathcal{E}_{j}[\cdot]$ on $\rho$ can be equivalently computed as the action of its conjugate channel $\mathcal{E}^*[\cdot]$ on $O^2=\sum_{ij}Z_i Z_j$. For the $X$ channel, this gives
\begin{equation}
    \mathcal{E}^*[O^{2}]=(1-2p)^2\sum_{i\neq j}Z_{i}Z_{j}+L,
\end{equation}
and in this way we can recover Eq.~\eqref{eq:spinflip}.
As another example, we consider the $ZZ$ channel $\mathcal{E}[\cdot]=\prod_{j}\mathcal{E}_{j}[\cdot]$, which is given as $\mathcal{E}_{j}(\rho_{0})=(1-p)\rho_{0}+pZ_{j}Z_{j+1}\rho_{0}Z_{j}Z_{j+1}$. Similarly, as $\left[Z_{j}Z_{j+1},\prod_{i}X_{i}\right]=0$, the parity measurement is optimal at $\theta\rightarrow0$ and the QFI equals exactly 4 times the variance [see discussion around Eq.~\eqref{eq:resolution_X} and Eq.~\eqref{eq:upperbound}]. Importantly, the action of the conjugate channel $\mathcal{E}^{*}_{j}(\cdot)$ on $O^2=\sum_{ij}Z_{i}Z_{j}$ is trivial, simplifying Eq.~\eqref{eq:spinflip} to
\begin{equation}\label{eq:QFIeqQFI}
F_{Q}[\rho]=F_{Q}[\rho_0].
\end{equation}
Therefore, both the optimal measurement and the QFI remain unchanged for the $ZZ$ channel. Since we do not use any specific property of the critical state in deriving Eq.~\eqref{eq:QFIeqQFI}, it holds for the spin squeezed and GHZ state as well. Nevertheless, for the SS, the optimal measurement is solely the parity measurement as the collective spin yields
\begin{align*}
    \prod_{j}\mathcal{E}^{*}_{j}\left[S_{\theta}\right]&=(1-2p)^2S_{\theta},\\
    \prod_{j}\mathcal{E}^{*}_{j}\left[S_{\theta}^{2}\right]&=(1-2p)^{4}S_{\theta}^{2}+O(L), 
\end{align*}
which gives the SQL in $L$ for any $p>0$ due to non-zero $O(L)$.
\section{Qubit loss: sub-Heisenberg-limit}
\label{appendix:XXZ}
This appendix provides the technical details underlying our main conclusions about the consequences of qubit losses in quantum metrology. We focus first on the Ising universality class, both in the Ising spin chain and Rydberg atoms, and then we repeat the analysis in the XXZ spin chain, aka the Luttinger liquid.
\begin{figure}[h]
    \centering
    \includegraphics[width=8cm]{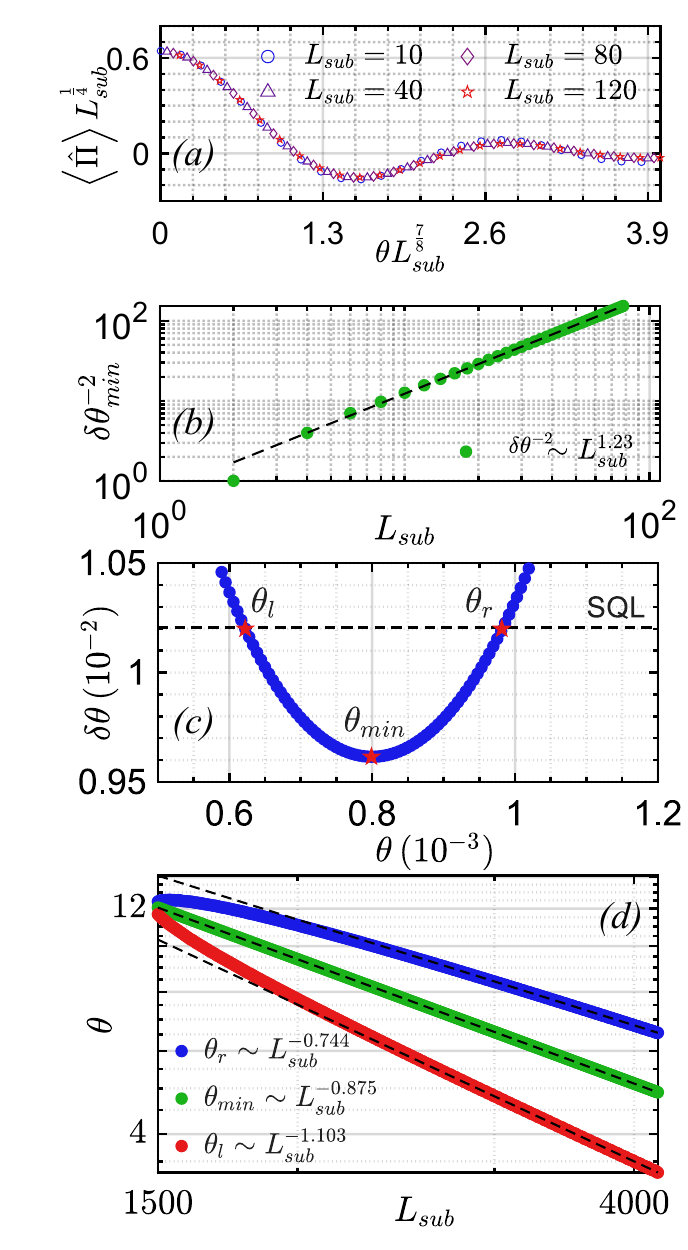}
    \caption{Parity measurement of the subsystem. (a). $\braket{\Pi_{\text{sub}}(\theta)}L_{\text{sub}}^{\frac{1}{4}}$ as a functions of $\theta L_{\text{sub}}^{\frac{7}{8}}$ for $L_{\text{sub}}=10,40,80,120$. (b). The optimal precision $\delta \theta_{min}$ as a function of sub-region length $L_{sub}$. (c).The phase precision of parity measurement over a sub region with $2400$ sites [SQL=$1/(2\sqrt{L_{\rm sub}})$]. (d). The scaling behavior of phase ranges $[\theta_{l},\theta_{r}]$ and the optimal phase $\theta_{min}$ that exhibit sub-SQL sensitivity. All results are obtained from iDMRG. The scaling behavior of $\delta\theta_{min}^{-2}$ in (b) is extracted at $\theta=\theta_{min}$ indicated in (c).}
    \label{fig:Parity_subsystem}
\end{figure}
\subsection{Ising universality class}
We start the analysis by computing $\delta\theta $ in Eq.~\eqref{eq:meansquare} using $\Pi_{\rm sub}=\prod_{j=1}^{L_{\rm sub}}X_j$.
The parity operator after the unitary rotation within the subsystem becomes  $\Pi_{\rm sub}(\theta)=U(\theta)\Pi_{\rm sub} U(\theta)^{\dagger}=\Pi_{\rm sub}U(\theta)^{\dagger 2}$. At criticality, the Ising model exhibits Kramers-Wannier duality, under which the spin field maps to the disorder field $\mu_j = \prod_{l \leq j} X_l$. This allows us to express the expectation value of $\Pi_{\rm sub}(\theta)$ as a two-point correlator in the dual picture

\begin{equation}
    \braket{\Pi_{\rm sub}(\theta)}_{c}=\braket{\mu_{0}\mu_{L_{\rm sub}}e^{-2i\theta\sum_{j=1}^{L_{\rm sub}-1}Z_{j}}}_{c}.\label{eq: parity expect}
\end{equation}
Expanding the exponential in powers of $\theta$, and noting that only even powers survive by symmetry, the coefficient of each term involves evaluating a $(2n+2)$-point connected correlator of disorder and spin operators
\begin{equation}
    \sum_{\vec{j}_{2n}}\braket{\mu^{z}_{0}\mu^{z}_{L_{\rm sub}}Z_{j_{1}}Z_{j_{2}}\cdots Z_{j_{2n}}}_{c}= c_{2n}(L_{\rm sub})\  L_{\rm sub}^{2n-n/4-1/4},
\end{equation}
where $\vec{j}_{2n}=(j_{1},\cdots,j_{2n})$ is a vector with each entry $j_{k}$ taking value in $(0,L_{\rm sub})$. In the power-law $L_{\rm sub}^{2n-\frac{n}{4}-1/4}$, the first $2n$ comes from the sum over $j_{1},\cdots,j_{2n}$, the term $n/4$ comes from critical exponent of $Z$ operators $\Delta_{Z}=1/8$, and the contribution $1/4$ comes from the critical exponent of each disorder field, $\Delta_{\mu}=1/8$. The scaling with $L_{\rm sub}$ has been extracted by a simple power-counting, since we that the coefficients $c_{2n}(L_{\rm sub})$ only show a subleading dependence on $L_{\rm sub}$ which does not affect the leading scaling behavior.

We therefore expect Eq.~\eqref{eq: parity expect} takes the following form
\begin{equation}
 \braket{\Pi_{\rm sub}(\theta)}_{c}=L_{\rm sub}^{-1/4} \sum_{n=0}^{+\infty}\frac{c_{2n}(L_{\rm sub})}{(2n)!}(-4\theta^2 L_{\rm sub}^{7/4})^n.\label{eq: parity theta}
\end{equation}
For $\theta = 0$, the expectation simply reduces to $\langle \mu_0 \mu_{L_{\rm sub}} \rangle_c = c_0 L_{\rm sub}^{-1/4}$. This immediately shows that $\theta=0$ no longer has the highest precision as in the ideal situation, without qubit loss.

Therefore, we first estimate the range of $\theta$ where at least the SQL, $\delta\theta \sim 1/\sqrt{L_{\rm sub}}$, is satisfied. For $|\theta| < L_{\rm sub}^{-7/8}$, higher-order terms in the series.~\eqref{eq: parity theta} are negligible, and the expansion can be truncated after the quadratic term
\begin{equation}
\partial_{\theta}\braket{\Pi_{\rm sub}({\theta})}_{c}\approx -4c_{2}L_{\rm sub}^{3/2}\theta. 
\end{equation}
Requiring $\left|\partial_{\theta}\braket{\Pi_{\rm sub}({\theta})}_{c}\right|\geq \sqrt{L_{\rm sub}}$ yields $\left|\theta\right|\geq L_{\rm sub}^{-1}$ up to constant prefactors.

When $\theta \gtrsim L_{\rm sub}^{-7/8}$, the higher-order terms in the expansion of $\langle \Pi_{\rm sub}(\theta) \rangle_c$ become significant and cannot be neglected. In this regime, we consider the full expression:
\begin{equation}
\left|\partial_{\theta}\braket{\Pi_{\rm sub}({\theta})}_{c}\right|=\left|\theta^{-1}L_{\rm sub}^{-\frac{1}{4}}\sum_{n=1}^{+\infty}\frac{c_{2n}}{(2n-1)!}\left(-4\theta^{2}L_{\rm sub}^{7/4}\right)^{n}\right|.\label{eq:parital_Parity_sub}
\end{equation}

We now assume that the series inside the sum converges to an $O(1)$ function, which is consistent with the behavior of $\langle \Pi_{\rm sub}(\theta) \rangle_c L_{\rm sub}^{1/4}$ in Fig.~\ref{fig:Parity_subsystem}a), where all curves are obtained from iDMRG with no approximation: For increasing system sizes $L_{\rm sub} = 10, 40, 80, 120$, the rescaled quantity $\braket{\Pi} L^{1/4}_{\text{sub}}$ converges uniformly to an $O(1)$ function of $\theta L^{7/8}_{\rm sub}$, indicating the rapid convergence of the coefficients $c_{2n}(L_{\text{sub}})$ in Eq.~\eqref{eq: parity theta} as $L_{\text{sub}}$ increases. 
The scaling of Eq.~\eqref{eq:parital_Parity_sub} then follows directly as $\left|\partial_\theta \langle \Pi_{\rm sub}(\theta) \rangle_c\right| \sim L_{\rm sub}^{-1/4} |\theta^{-1}|$. Imposing the SQL condition, we then deduce the bound $|\theta| \lesssim L_{\rm sub}^{-3/4}$, up to an $O(1)$ prefactor. 

Within the range $L_{\mathrm{sub}}^{-1}<\theta< L_{\mathrm{sub}}^{-3/4}$, we want to find the parameter $\theta$ where the parity measurement has the highest sensitivity, i.e. maximal~\eqref{eq:meansquare}. Fig.~\ref{fig:Parity_subsystem}(a) shows the collapse of the rescaled expectation value
\begin{equation}
\langle \Pi_{\rm{sub}}(\theta)\rangle L_{\rm{sub}}^{1/4}
\end{equation}
onto an $\mathcal{O}(1)$ scaling function $f(\theta L_{\rm{sub}}^{7/8})$. The estimator~\eqref{eq:meansquare} reads
\begin{equation}
\delta\theta^{-2}
= L_{\rm{sub}}^{-1/2}
\frac{|\partial_{\theta} f|^{2}}{1 - L_{\rm{sub}}^{-1/2} f^{2}}.
\end{equation}
Since $f=\mathcal{O}(1)$, the denominator satisfies
$1 - L_{\rm{sub}}^{-1/2} f^{2} \approx 1$, allowing us to determine the angle $\theta_{\rm min}$ where the sensitivity is maximal. This occurs at the point
\begin{equation}\label{eq:deltamin}
\theta_{\rm min} = \theta_0 L_{\rm{sub}}^{-7/8},
\end{equation}
with $\theta_0$ defined by the inflection condition $f''(\theta_0)=0$. Evaluating Eq.~\eqref{eq:meansquare}, the sensitivity at this optimal angle gives
\begin{equation}
\delta\theta = \left[f'(\theta_0)\right]^{-1/2} L_{\rm{sub}}^{-5/8},
\end{equation}
where $f'(\theta_0)$ denotes the derivative of $f$ at $\theta_0$. Thus, the optimal sensitivity of the parity measurement improves over the standard quantum limit by a factor $L_{\rm{sub}}^{-1/8}$.

To summarize, our analysis is valid within the range of angles $\theta$ for which the phase uncertainty $\delta\theta$ remains below the SQL, specifically in the interval $\theta_l < \theta < \theta_r$, where $\theta_l \sim L_{\rm sub}^{-1}$ and $\theta_r \sim L_{\rm sub}^{-3/4}$. We also identify an intermediate optimal point $\theta_{min} \sim L_{\rm sub}^{-7/8}$, where $\delta\theta$ takes the most accurate value as quantified by the scaling with subsystem size. Besides the above analysis, we numerically evaluate $\delta\theta$ as a function of $\theta$ for different values of $L_{\rm sub}$: We show in Fig.~\ref{fig:Parity_subsystem}b) the minimal phase uncertainty for various values of $L_{\rm sub}$. Then we extract the interval where the uncertainty $\delta\theta$ is below the SQL [Fig.~\ref{fig:Parity_subsystem}c) for $L_{\text{sub}}=2400$], and collect the corresponding $\theta_{l},\theta_{min},\theta_{r}$ as functions of $L_{\rm sub}$ in Fig.~\ref{fig:Parity_subsystem}d). From these plots, we verify that $\theta_{\min}\sim L_{\rm sub}^{-7/8}$, which satisfies Eq.~\eqref{eq:deltamin},
and $\delta \theta_{\min}\sim L_{\rm sub}^{-5/8}$. The scaling behavior of $\theta_l$, $\theta_r$ is also supported by numerical data in Fig.~\ref{fig:Parity_subsystem}d).

A natural question that one might ask at this point is whether the same conclusion holds when qubit losses destroy $\mathbb{Z}_2$ spatial symmetries, like the translational invariance discussed for the Hamiltonian \eqref{eq:RydbergH}. A microscopic realization of the disorder operator in this case is given by
\cite{Kevin2021}
\begin{align}
    \mu_{j}&=\cdots S_{j-\frac{5}{2}}S_{j-\frac{3}{2}}S_{j-\frac{1}{2}}\zeta_{j},
\end{align}
with  $S_{j+\frac{1}{2}}\ket{n_{j}n_{j+1}}=\ket{n_{j+1}n_{j}}$ the swap operator and $\zeta_{j}=\ket{0_{j}}\left(\sqrt{1-\braket{n}}\bra{0_{j}}-\sqrt{\braket{n}}\bra{1_{j}}\right)$. The string of SWAP operators effectively moves the degree of freedom at site 
$j$ to the left edge of the chain, mimicking the nonlocal string that appears in the Kramers–Wannier duality. The final operator $\zeta_j$ projects site $j$ onto the “typical’’ local wavefunction $\ket{\psi_j}=\sqrt{1-\braket{n}}\ket{0_j}-\sqrt{\braket{n}}\ket{1_j}$ favored by the Rabi drive, characterized by the average occupation $\braket{n}$ and the correct sign structure for $\Omega
>0$, and then parks the site in the empty state $\ket{0}$, thereby disentangling it from the rest of the chain. Together, these operations identify whether a domain-wall endpoint is present at site 
$j$, reproducing the structure of the Ising disorder operator in the microscopic Rydberg Hilbert space.

As we did for the Ising model, we can define the measurement operator restricted to the subsystem as $\Pi_{\text{sub}}=\mu^{\dagger}_{0}\mu_{L_{\text{sub}}}$ and the imprinting unitary $U(\theta)=e^{iO_{\text{sub}}\theta}$ with $O_{\text{sub}}=\sum_{0<j<L_{\text{sub}}}\sigma_{j}$. Since $\{\sum_{j}\sigma_{j},\ \mu_{0}^{\dagger}\mu_{L_{\text{sub}}}\}=0$, the expectation value of the measurement operator after the unitary rotation becomes
\begin{equation}
    \braket{\Pi_{\text{sub}}(\theta)}=\braket{\mu_{0}^{\dagger}\mu_{L_{\text{sub}}}e^{-2i\theta\sum_{0<j<L_{\text{sub}}}\sigma_{j}}}.
\end{equation}
At the field theory level, this is exactly the same equation as \eqref{eq: parity expect}, so it would yield the same conclusion for the scaling of precision as $\delta\theta_{\text{min}}$ as $ L_{\text{sub}}^{-5/8}$.
This result is quite remarkable because, even though qubit losses break the translational invariance of the system, we can still find an advantage to the SQL. 

\subsection{Tricritical Ising universality class}

\begin{table}
\begin{center}
\begin{tabular}{|c|c|c|}
    \hline
     & \textbf{TFIM} & \textbf{BCM}\\ \hline
    $\delta\theta_{min}$ & $O(L_{\text{sub}}^{-\frac{5}{8}})$& $O(L_{\text{sub}}^{-\frac{31}{40}})$\\ \hline 
    $\theta_{l}$ & $O(L_{\text{sub}}^{-1})$
    &$O(L_{\text{sub}}^{-\frac{6}{5}})$\\ \hline 
    $\theta_{min}$ & $O(L_{\text{sub}}^{-\frac{7}{8}})$ & $O(L_{\text{sub}}^{-\frac{37}{40}})$\\ \hline
    $\theta_r$&$O(L_{\text{sub}}^{-\frac{3}{4}})$&$O(L_{\text{sub}}^{-\frac{13}{20}})$\\ \hline 
\end{tabular}
\end{center}
    \caption{Scaling behavior of the phase precision and the sub-SQL window for a finite region $L_{\textbf{sub}}$ of the transverse-field Ising model and the Blume-Capel model. }
    \label{tab:qubit_loss}
\end{table}

\begin{figure}[h]
    \centering
    \includegraphics[width=7cm]{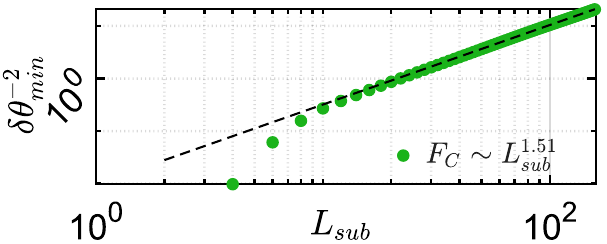}
    \caption{The optimal precision $\delta \theta_{min}$ from the subsystem parity measurement as a function of sub-region length $L_{sub}$ for the Blume-Capel spin-$1$ model at the tricritical point.}
    \label{fig:Parity_subsystem_trising}
\end{figure}

The argument for the subsystem metrology for the transverse-field Ising model can be naturally generalized to other critical systems via Kramer-Wannier (KW) duality or Jordan-Wigner (JW) transformation. Let us now consider a different critical model---the Blume-Capel spin-$1$ model (BCM) described by
\begin{equation}
    H_{BC}=- J \sum_{i} S_i^{Z} S_{i+1}^{Z}
+ D \sum_{i} (S_i^{Z})^{2}
- \Gamma \sum_{i} S_i^{X},
\end{equation}
where $S_i^{X,Z}$ are spin-$1$ angular momentum operators at site $i$. The tricritical point for this model is numerically found at $D/J=0.91021$ and $\Gamma/J=0.41558$, where there are two sets of charge and disorder fields with different scaling dimensions $\Delta_{\sigma}=\Delta_{\mu}=3/40$ and $\Delta_{\sigma'}=\Delta_{\mu'}=7/8$. We focus on the first set with with scaling dimension $3/40$ that dominates the long range behavior. The lattice realizations of relevant primary fields are $\sigma_i\sim S_{i}^{Z}$ and $\mu_i\sim\prod_{j<i}P_j^{X}$ with 
\begin{equation}
    S_{j}^Z=\begin{pmatrix}
1 & 0 & 0 \\
0 & 0 & 0 \\
0 & 0 & -1
\end{pmatrix},\ \ P_{j}^X=\begin{pmatrix}
0 & 0 & 1 \\
0 & 1 & 0 \\
1 & 0 & 0
\end{pmatrix}\label{eq: tricritical P operator}
\end{equation}
represented in the $Z$ basis. The subsystem parity $\Pi_{\text{sub}}$[Eq.~\eqref{Parity_sub}] is replaced by the product of two disorder fields $\mu_0\mu_{L_{\text{sub}}}$, whose lattice version is identified as
\begin{equation}
\Pi_{\text{sub}}=\prod_{i=1}^{L_{\text{sub}}}P_{i}^{X}.
\end{equation}
After rotation by $e^{i\theta\sum_{j=1}^{L_{\text{sub}}}S_j^{Z}}$, a similar expression to Eq.~\eqref{eq: parity expect} can be obtained as 
\begin{equation}
    \braket{\Pi_{\rm sub}(\theta)}_{c}=\braket{\mu_{0}\mu_{L_{\rm sub}}e^{-2i\theta\sum_{j=1}^{L_{\rm sub}-1}S_{j}^{Z}}}_{c}.
\end{equation}
 We can now repeat the calculation from Eq.~\eqref{eq: parity expect} to Eq.~\eqref{eq:parital_Parity_sub} and find the scaling of $\delta\theta_{min}$ and $\theta_{l,r,min}$. We summarize our results in Table~\ref{tab:qubit_loss}. Crucially, as the scaling dimension for the tricritical Ising universality class is much smaller than that for the Ising universality class, the minimal precision $\delta\theta_{min}$ converges much closer to the Heisenberg limit $L_{\text{sub}}^{-1}$ [Fig.~\ref{fig:Parity_subsystem_trising}]. Crucially, the phase window $\left[\theta_l,\theta_r\right]\rightarrow\left[O(L_{\text{sub}}^{-\frac{6}{5}}), O(L_{\text{sub}}^{-\frac{13}{20}})\right]$ becomes much wider than the precision $\delta\theta\sim L^{-\frac{31}{40}}_{\text{sub}}$ compared to the Ising criticality, indicating a practically meaningful sensing advantage.

Before ending, we also note that the identification of the subsystem parity can also be obtained for fermion systems, with the aid of the JW transformation: the subsystem parity $\Pi_{\text{sub}}$ [Eq.~\eqref{Parity_sub}] can be chosen as the string in the JW transform
\begin{equation}
    \gamma_{j,1}= X_{j}\prod_{i<j}(-Z_{i}),\quad \gamma_{j,2}=Y_{j}\prod_{i<j}(-Z_{i}),\label{eq:JW}
\end{equation}
where $\gamma_{j,1},\gamma_{j,2}$ are Majorana fermions and the subsystem parity is identified as 
\begin{equation}
    \Pi_{\text{sub}}^{(\alpha, \beta)}=i\gamma_{0,\alpha}\gamma_{L_{\text{sub}},\beta}.
\end{equation}
In total, there are $4$ possible choices of $\Pi_{\text{sub}}^{(\alpha, \beta)}$ with $\alpha, \beta=1,2$. However, some of them vanish due to symmetries of the system: the time-reversal symmetry $\mathcal{T}$ flips $\Pi^{(\alpha, \beta)}_{L_{\text{sub}}}$ for $i= j$, leading to vanishing two-point correlation $\braket{\Pi^{(\alpha, \beta)}_{L_{\text{sub}}}}=0$. Another challenge for measuring such subsystem parity is from the fast decaying of fermion correlations. Therefore, although subsystem QFI could exhibit super-SQL scaling, the fermion correlation might hinder it. For example, consider the XXZ model where an improvement in the QFI comes from tuning $K\rightarrow\infty$, such that $\braket{X_{0}X_{x}}=\braket{Y_{0}Y_{x}}\sim x^{-\frac{1}{2K}}$ becomes perfectly long-range correlated. However, this unavoidably compromises the fermion correlations $\braket{\gamma_{0,i}\gamma_{x,i}}\sim x^{-2K}$. It is then interesting to ask what the right subsystem operator is, whose precision beats the SQL. We leave such a discussion for future work.

\end{document}